\documentclass[twocolumn]{aastex631}
\usepackage{multirow}
\usepackage{color}
\usepackage{float}
\usepackage{tabularx}
\makeatother
\makeatletter
\usepackage{comment}
\usepackage{natbib}
\usepackage{amsmath}
\usepackage{booktabs}
\usepackage{array}
\usepackage{hyperref}
\hypersetup{colorlinks=true, citecolor=blue, filecolor=magenta,
 urlcolor=cyan, linkcolor=blue}

\shortauthors{Long et al.}

\begin{document}

\title{Efficient NIRCam Selection of Quiescent Galaxies at $3 < z < 6$ in CEERS}

\author[0000-0002-7530-8857]{Arianna S. Long}\altaffiliation{NASA Hubble Fellow; E-mail: arianna.sage.long@gmail.com}
\affiliation{Department of Astronomy, The University of Texas at Austin, Austin, TX, USA}

\author[0000-0002-0243-6575]{Jacqueline Antwi-Danso}
\affiliation{George P. and Cynthia Woods Mitchell Institute for Fundamental Physics and Astronomy, Texas A\&M University, College Station, TX, USA}
\affiliation{Department of Physics and Astronomy, Texas A\&M University, 4242 TAMU, College Station, TX, USA}

\author[0000-0003-3216-7190]{Erini L. Lambrides}
\altaffiliation{NASA Postdoctoral Fellow}
\affiliation{NASA-Goddard Space Flight Center, Code 662, Greenbelt, MD, 20771, USA}

\author[0000-0001-7964-5933]{Christopher C. Lovell}
\affiliation{Institute of Cosmology and Gravitation, University of Portsmouth, Burnaby Road, Portsmouth PO1 3FX, UK}
\affiliation{Centre for Astrophysics Research, School of Physics, Engineering \& Computer Science, University of Hertfordshire, Hatfield AL10 9AB, UK}

\author[0000-0002-6219-5558]{Alexander de la Vega}
\affiliation{Department of Physics and Astronomy, University of California, 900 University Ave, Riverside, CA 92521, USA}

\author[0000-0001-6477-4011]{Francesco Valentino}
\affiliation{Cosmic Dawn Center (DAWN), Denmark}
\affiliation{Niels Bohr Institute, University of Copenhagen, Jagtvej 128, DK-2200 Copenhagen N, Denmark}
\affiliation{European Southern Observatory, Karl-Schwarzschild-Str. 2, D-85748 Garching bei Munchen, Germany}

\author[0000-0002-7051-1100]{Jorge A. Zavala}
\affiliation{National Astronomical Observatory of Japan, 2-21-1 Osawa, Mitaka, Tokyo 181-8588, Japan}

\author[0000-0002-0930-6466]{Caitlin M. Casey}
\affiliation{Department of Astronomy, The University of Texas at Austin, Austin, TX, USA}

\author[0000-0003-3903-6935]{Stephen M.~Wilkins} %
\affiliation{Astronomy Centre, University of Sussex, Falmer, Brighton BN1 9QH, UK}
\affiliation{Institute of Space Sciences and Astronomy, University of Malta, Msida MSD 2080, Malta}

\author[0000-0003-3466-035X]{{L. Y. Aaron} {Yung}}
\altaffiliation{NASA Postdoctoral Fellow}
\affiliation{Astrophysics Science Division, NASA Goddard Space Flight Center, 8800 Greenbelt Rd, Greenbelt, MD 20771, USA}

\author[0000-0002-7959-8783]{Pablo Arrabal Haro}
\affiliation{NSF's National Optical-Infrared Astronomy Research Laboratory, 950 N. Cherry Ave., Tucson, AZ 85719, USA}

\author[0000-0002-9921-9218]{Micaela B. Bagley}
\affiliation{Department of Astronomy, The University of Texas at Austin, Austin, TX, USA}

\author[0000-0003-0492-4924]{Laura Bisigello}
\affiliation{Dipartimento di Fisica e Astronomia "G.Galilei", Universit\'a di Padova, Via Marzolo 8, I-35131 Padova, Italy}
\affiliation{INAF--Osservatorio Astronomico di Padova, Vicolo dell'Osservatorio 5, I-35122, Padova, Italy}

\author[0000-0003-4922-0613]{Katherine Chworowsky}\altaffiliation{NSF Graduate Fellow}
\affiliation{Department of Astronomy, The University of Texas at Austin, Austin, TX, USA}

\author[0000-0003-1371-6019]{M. C. Cooper}
\affiliation{Department of Physics \& Astronomy, University of California, Irvine, 4129 Reines Hall, Irvine, CA 92697, USA}

\author[0000-0003-3881-1397]{Olivia R. Cooper}\altaffiliation{NSF Graduate Fellow}
\affiliation{Department of Astronomy, The University of Texas at Austin, Austin, TX, USA}

\author[0000-0002-3892-0190]{Asantha R. Cooray}
\affiliation{Department of Physics \& Astronomy, University of California, Irvine, 4129 Reines Hall, Irvine, CA 92697, USA}

\author[0000-0002-5009-512X]{Darren Croton}
\affiliation{Centre for Astrophysics \& Supercomputing, Swinburne University of Technology, Hawthorn, VIC 3122, Australia}
\affiliation{ARC Centre of Excellence for All Sky Astrophysics in 3 Dimensions (ASTRO 3D)}

\author[0000-0001-5414-5131]{Mark Dickinson}
\affiliation{NSF's National Optical-Infrared Astronomy Research Laboratory, 950 N. Cherry Ave., Tucson, AZ 85719, USA}

\author[0000-0001-8519-1130]{Steven L. Finkelstein}
\affiliation{Department of Astronomy, The University of Texas at Austin, Austin, TX, USA}

\author[0000-0002-3560-8599]{Maximilien Franco}
\affiliation{Department of Astronomy, The University of Texas at Austin, Austin, TX, USA}

\author[0000-0003-4196-5960]{Katriona M. L. Gould}
\affiliation{Cosmic Dawn Center (DAWN), Denmark}
\affiliation{Niels Bohr Institute, University of Copenhagen, Jagtvej 128, DK-2200 Copenhagen N, Denmark}

\author[0000-0002-3301-3321]{Michaela Hirschmann}
\affiliation{Institute of Physics, Laboratory of Galaxy Evolution, Ecole Polytechnique Fédérale de Lausanne (EPFL), Observatoire de Sauverny, 1290 Versoix, Switzerland}

\author[0000-0001-6251-4988]{Taylor A. Hutchison}
\altaffiliation{NASA Postdoctoral Fellow}
\affiliation{Astrophysics Science Division, NASA Goddard Space Flight Center, 8800 Greenbelt Rd, Greenbelt, MD 20771, USA}

\author[0000-0001-9187-3605]{Jeyhan S. Kartaltepe}
\affiliation{Laboratory for Multiwavelength Astrophysics, School of Physics and Astronomy, Rochester Institute of Technology, 84 Lomb Memorial Drive, Rochester, NY 14623, USA}

\author[0000-0002-8360-3880]{Dale D. Kocevski}
\affiliation{Department of Physics and Astronomy, Colby College, Waterville, ME 04901, USA}

\author[0000-0002-6610-2048]{Anton M. Koekemoer}
\affiliation{Space Telescope Science Institute, 3700 San Martin Dr.,
Baltimore, MD 21218, USA}

\author[0000-0003-1581-7825]{Ray A. Lucas}
\affiliation{Space Telescope Science Institute, 3700 San Martin Drive, Baltimore, MD 21218, USA}

\author[0000-0002-6149-8178]{Jed McKinney}
\affiliation{Department of Astronomy, The University of Texas at Austin, Austin, TX, USA}

\author[0000-0001-7503-8482]{Casey Papovich}
\affiliation{Department of Physics and Astronomy, Texas A\&M University, College Station, TX, 77843-4242 USA}
\affiliation{George P.\ and Cynthia Woods Mitchell Institute for Fundamental Physics and Astronomy, Texas A\&M University, College Station, TX, 77843-4242 USA}

\author[0000-0003-4528-5639]{Pablo G. P\'erez-Gonz\'alez}
\affiliation{Centro de Astrobiolog\'{\i}a (CAB), CSIC-INTA, Ctra. de Ajalvir km 4, Torrej\'on de Ardoz, E-28850, Madrid, Spain}

\author[0000-0003-3382-5941]{Nor Pirzkal}
\affiliation{ESA/AURA Space Telescope Science Institute}

\author[0000-0002-9334-8705]{Paola Santini}
\affiliation{INAF - Osservatorio Astronomico di Roma, via di Frascati 33, 00078 Monte Porzio Catone, Italy}

% \author[0000-0003-3216-7190]{Erini L. Lambrides} \altaffiliation{NPP Fellow}
% \affiliation{NASA Goddard Space Flight Center, 8800 Greenbelt Road Greenbelt, MD 20771, USA}

% \author[0000-0003-3881-1397]{Olivia R. Cooper}\altaffiliation{NSF Graduate Fellow}
% \affiliation{The University of Texas at Austin, Department of Astronomy, Austin, TX, United States}

\begin{abstract}

Substantial populations of massive quiescent galaxies at $z\ge3$ challenge our understanding of rapid galaxy growth and quenching over short timescales. In order to piece together this evolutionary puzzle, more statistical samples of these objects are required. Established techniques for identifying massive quiescent galaxies are increasingly inefficient and unconstrained at $z>3$. As a result, studies report that as much as 70\% of quiescent galaxies at $z>3$ may be missed from existing surveys. In this work, we propose a new empirical color selection technique designed to select massive quiescent galaxies at $3\lesssim z \lesssim 6$ using JWST NIRCam imaging data. We use empirically-constrained galaxy SED templates to define a region in the $F277W-F444W$ vs. $F150W-F277W$ color plane that captures quiescent galaxies at $z>3$. We apply this color selection criteria to the Cosmic Evolution Early Release Science (CEERS) Survey and identify 44 candidate $z\gtrsim3$ quiescent galaxies. Over half of these sources are newly discovered and, on average, exhibit specific star formation rates of post-starburst galaxies. We derive volume density estimates of $n\sim1-4\times10^{-5}$\,Mpc$^{-3}$ at $3<z<5$, finding excellent agreement with existing reports on similar populations in the CEERS field. Thanks to NIRCam's wavelength coverage and sensitivity, this technique provides an efficient tool to search for large samples of these rare galaxies. 

\end{abstract}


\section{Introduction} \label{sec:intro}

One of the most puzzling discoveries in galaxy evolution of the decade is that substantial populations of massive galaxies (M$_*$\,$\gtrsim10^{10.5}$\,M$_\odot$) ceased forming stars as early as two billion years after the Big Bang \cite[$z$\,$>3$, ][]{toft14, Glazebrook2017, Merlin2019, Shahidi2020, Valentino2020, Forrest2020, Carnall2023}. Many of the best available cosmological models struggle to produce adequate populations -- if any at all -- of massive quiescent galaxies at $z>3$ (\citealt{Steinhardt2016, Schreiber2018, Cecchi2019}, see also \textsc{eagle} in \citealt{Merlin2019} and \textsc{flares} in \citealt{Lovell2022}), highlighting our incomplete understanding of the physics required to quench these behemoths over the short periods of time available in the early cosmos. Recent discoveries of surprisingly massive galaxies at $z>6$ apply even further pressure on our understanding of cosmology and galaxy quenching physics \citep{Labbe2022, Robertson2022, Boylan-Kolchin2022, Lovell2023} as these galaxies must form and quench in $\lesssim$\,1\,Gyr to match observations at later times. Therefore, identifying and characterizing large samples of early massive quiescent galaxies is fundamental to testing our theories on the construction of the first massive galaxies.

% quiescent galaxies at $z>3$ are exceedingly rare \citep[n\,$\sim10^{-5}-10^{-6}$\,Mpc$^{-3}$][]{}, meaning that wide-field surveys with deep, multi-band coverage are critical to detecting statistically significant samples of these objects. Quiescent galaxies at $z>3$ are a rare population, with number densities on the order of n\,$\sim10^{-5}-10^{-6}$\,Mpc$^3$ \citep{Girelli2019, Merlin2019, Santini2019, Shahidi2020}.\citep[n\,$\sim10^{-5}-10^{-6}$\,Mpc$^{-3}$][]{Girelli2019, Santini2019, Shahidi2020, Valentino2020, Long2022, Carnall2023}

% Understanding how this population formed and quenched in a few Gyr is critical to constraining the impact of a variety of important astrophysical processes in galaxy evolution including the large scale distribution of matter, extreme star formation and/or active galactic nuclei (AGN) feedback, dark matter halo growth, as well as gas accretion, recycling, and consumption cycles. \textbf{insert citations for these.} Maybe also reframe / add some motivation on why it's hard to identify them at $3<z<5$. 

% Reproducing the observed number density of such galaxies at all epochs is a compelling concern of current galaxy evolution models in cosmological simulations, as the relative importance of the quenching mechanisms is not yet clear (Man & Belli 2018). NOT MY SENTENCE

In order to understand the formation of massive quiescent galaxies in the $z>3$ cosmos, we must first securely identify statistical samples of these extreme objects. Quiescent galaxies at $z>3$ are exceedingly rare \citep[n\,$\sim10^{-5}-10^{-6}$\,Mpc$^{-3}$, ][]{Girelli2019, Santini2019, Shahidi2020, Valentino2020, Long2022, Carnall2023}, making their identification a strenuous process. Over the last two decades, identification techniques for quiescent galaxies were established primarily through rest-frame color-color diagrams and/or their relative position to the coeval star-forming main sequence (i.e. specific star formation rates, sSFRs). At $z>3$, both of these methods require generating galaxy spectral energy distributions (SEDs) and corresponding photometric redshifts, which are heavily entangled with the limitations and assumptions that go into galaxy SED modeling \citep[e.g. the shapes of star formation histories,][]{Merlin2018, Shahidi2020}. The majority of color selection techniques for these objects use a galaxy's rest-frame $J$-band magnitude to distinguish between dust-reddened star forming galaxies and quiescent galaxies \citep[see e.g. the $UVJ$ diagram, ][]{Labbe2005, Williams2009, Brammer2011}. However, at $z>3$ the rest-frame $J$-band redshifts to wavelengths $> 5$\,$\mu$m, falling into a spectral window that typically has no direct observational constraints (and if there are constraints through e.g. \textit{Spitzer} IRAC data, it is often too shallow to reliably detect these galaxies). In these cases, studies instead interpolate the rest-frame $J$-band magnitude from an unconstrained portion of the galaxy's SED, and are therefore model dependent. The MIRI instrument on JWST, with imaging coverage between $\lambda \approx 6-25$\,$\mu$m, could aid in rest-frame $J$-band measurements at $z>3$, however it has a much smaller on-sky footprint for the majority of JWST Cycle 1 legacy surveys currently underway (e.g. only a single band of MIRI imaging is scheduled to cover $\sim35$\% of the widest Cycle 1 JWST survey, COSMOS-Web, \citealt{Casey2022}). Given the rarity of $z>3$ quiescent galaxies, such a lack of coverage is a severe challenge for methods that employ rest-frame $J$-band magnitudes in high-$z$ quiescent galaxy searches.

% The \textit{UVJ} diagram relies on a bimodality in $UVJ$ color space between star forming and quiescent galaxies that persists out to $z\sim2$ \citep{Labbe2005, Williams2009, Brammer2011}, with some studies pushing this diagnostic out to as far as $z\sim4$ \citep{Whitaker2011, muzzin13, Straatman2014}. However, there are several major issues in extending the $UVJ$ method beyond $z=3$. 

Another major problem with current color selection methods is that they are tuned to the $z<2$ Universe, where there is a strong color bimodality in the quiescent versus star-forming galaxy population \citep{Labbe2005, Williams2009, Brammer2011}. At $z>3$, this bimodality is significantly less distinct \citep{Whitaker2011, muzzin13, Straatman2016} as many quiescent galaxies at high-$z$ exhibit young ages ($<500$\,Myr), making them appear bluer than typical quiescent galaxies at $z<3$ \citep{Forrest2020, D'Eugenio2020, Stevans2021, Perez-Gonzalez2022}. This is likely because not enough time has passed to allow for their stellar populations to fully age and redden to match the colors of quiescent galaxies at $z<2$ \citep{Merlin2018, Lovell2022}. Depending on the method, this can result in catastrophic losses on sample completeness, whereas much as $\sim$\,70\% of quiescent galaxies at these epochs could be entirely missed \citep{deshmukh18, Merlin2018, Valentino2020, Lovell2022}. Furthermore, the rest-frame colors of galaxies with heavily dust-reddened stellar spectra can masquerade as quiescent galaxy colors in the $z>1$ cosmos \citep{Hwang2021, deshmukh18, Martis2019}; and, depending on the epoch, dust-obscured galaxies may have similar population densities \citep{toft14, simpson14, Long2022}. These degeneracies yield significant rates ($10-40\%$) of false positives in the hunt for quiescent galaxies at high-$z$.

% \textbf{Insert note on language -- quiescent, passive, dormant, and how that relates to epoch of observation}?

\begin{figure*}[!ht]
\begin{center}
\includegraphics[trim=0.cm 1cm 0.1cm 0cm, width=0.3\textwidth]{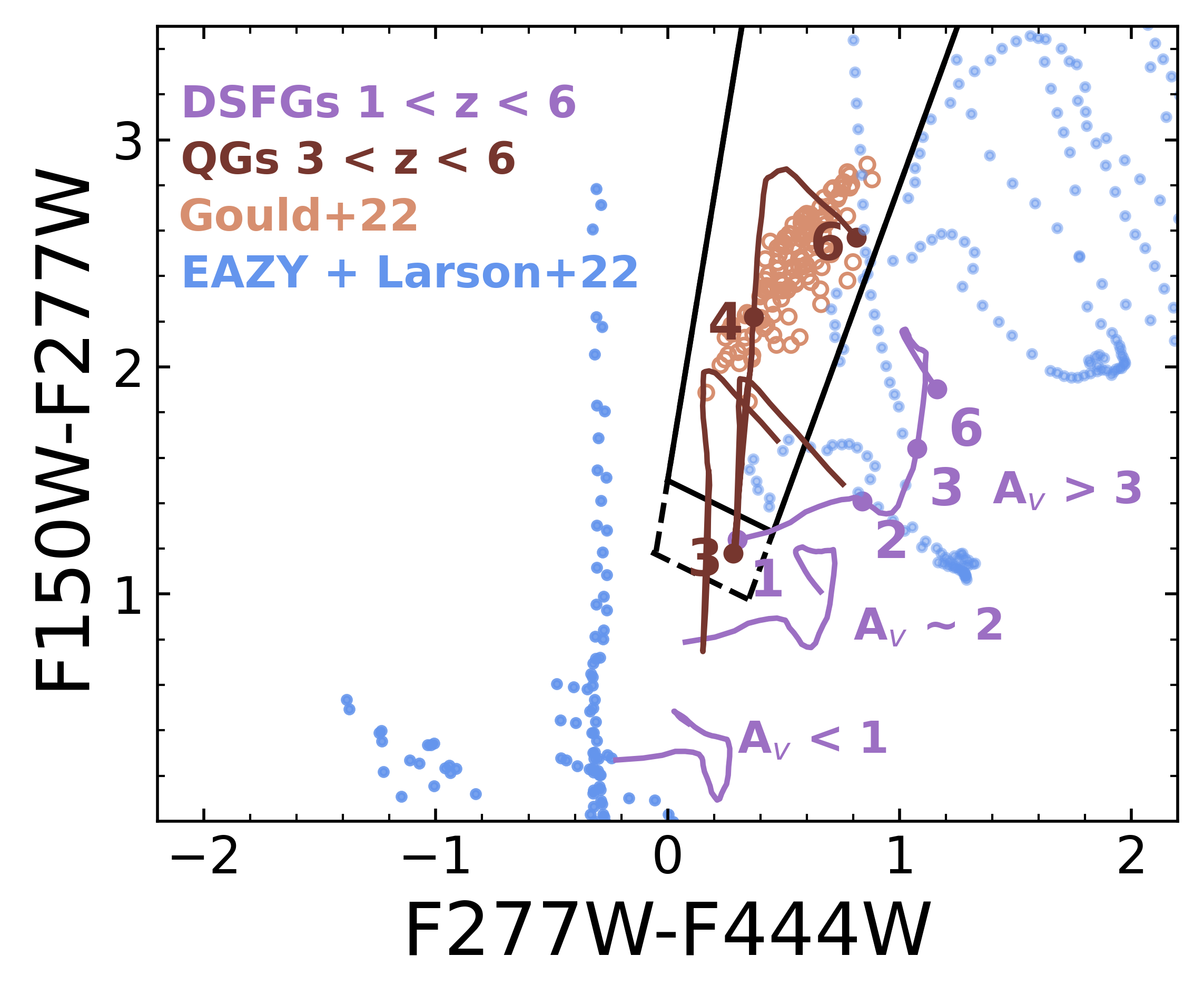}
\includegraphics[trim=0.cm 1cm 0.25cm 0cm, width=0.34\textwidth]{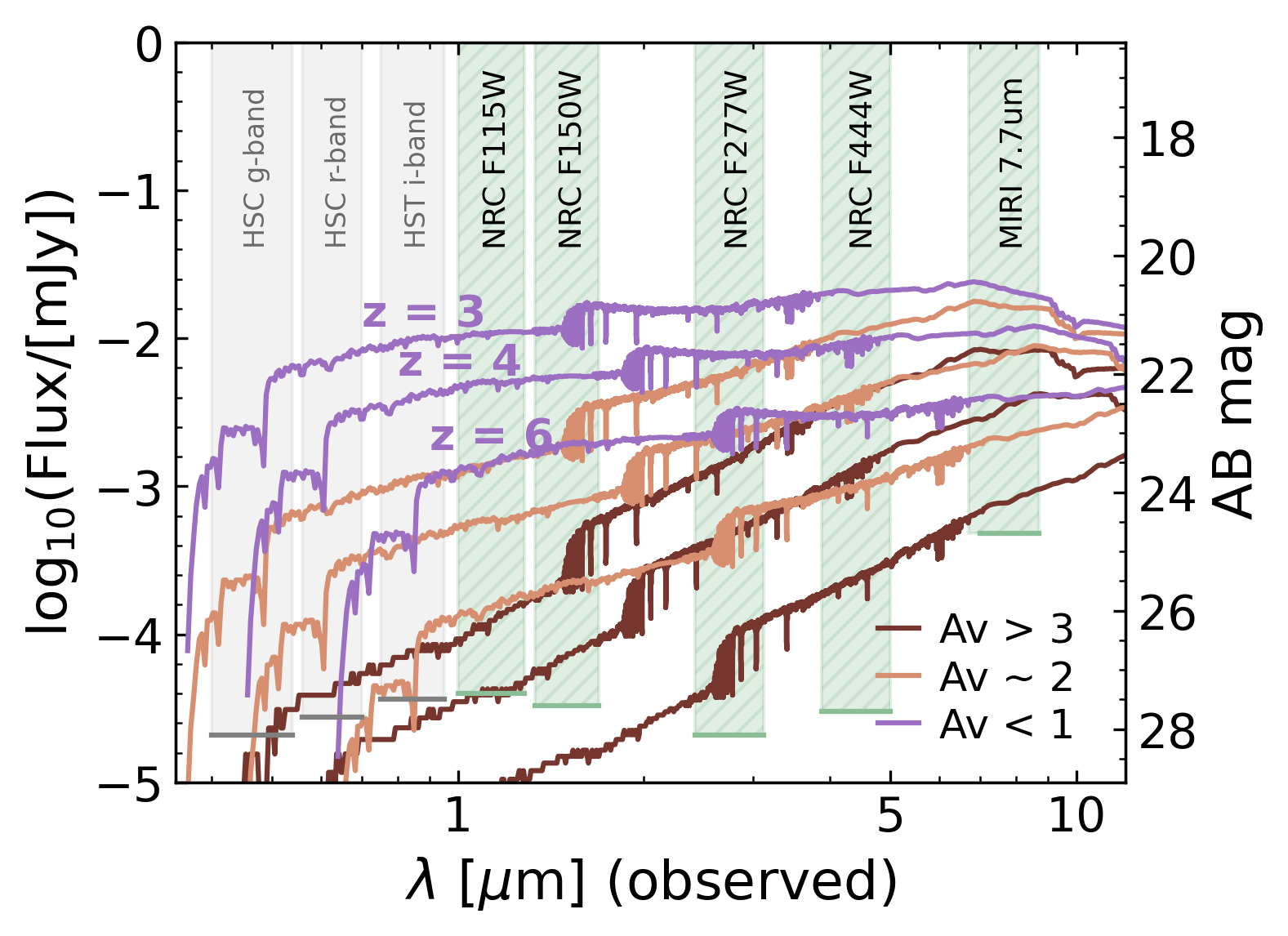}
\includegraphics[trim=0.cm 1cm 0.25cm 0cm, width=0.34\textwidth]{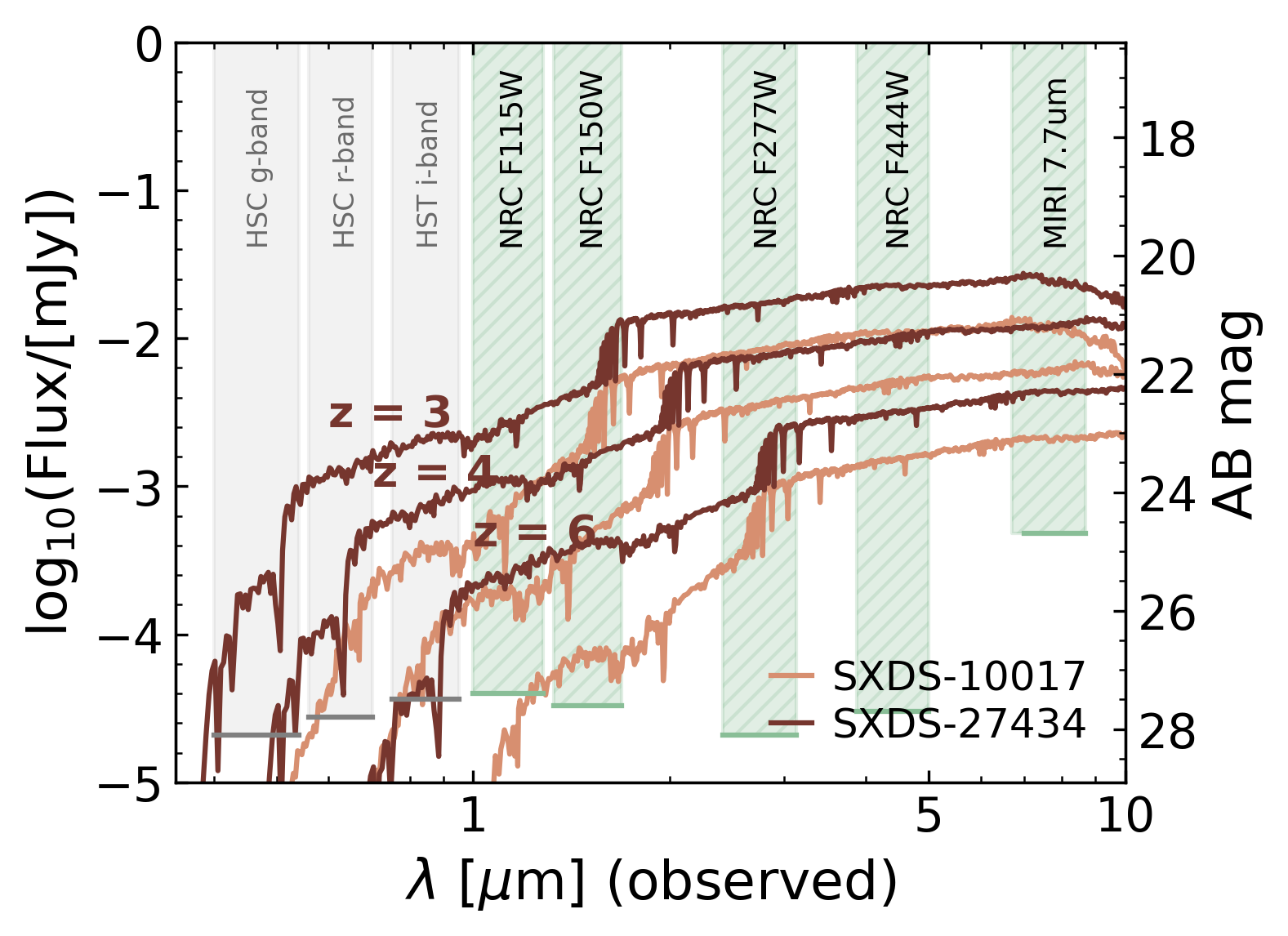}
\end{center}
\caption{\textbf{Left:} $F277W-F444W$ vs. $F150W-F277W$ colors for well-constrained galaxy SED templates. The black line marks the ``short'' wedge boundaries, while the dashed line marks the ``long'' wedge extension. The former is less complete but also has fewer contaminants, while the latter captures most quiescent galaxies at $3<z<6$ but with higher rates of contamination. We also show DSFG SEDs redshifted from $z=1$ to 6 in purple \citep[][also shown in the middle panel]{dacunha15}, quiescent galaxies at $z\sim4$ redshifted from $z=3$ to 6 in red \cite[][also shown in the rightmost panel]{Valentino2020}, and emission line galaxy SEDs redshifted from $z=3$ to 20 in blue \citep{Larson2022}. \textbf{Middle:} Observed-frame galaxy spectral energy distributions redshifted to $z=3$, 4 and 6. As an example JWST survey that is ideal for this search, we over lay the 5$\sigma$ photometric limits of various wideband surveys that cover the COSMOS-Web field \citep{Casey2022}. purple, orange, and brown lines are the redshifted SEDs of $z\sim2-3$ ALESS submillimeter galaxies from \citet{dacunha15}, progressing from low V-band attenuation ($A_v < 1$), to average ($A_v \sim 2$), to highly attenuated ($A_v \ge 3$), respectively. \textbf{Right:} SEDs of spectroscopically confirmed $z\sim4$ quiescent galaxies from \citet{Valentino2020}, redshifted to $z=3$, 4, and 6. The brown SED corresponds to SXDS-10017, a more evolved quiescent galaxy at $z_\mathrm{spec} = 3.767$, and the orange SED corresponds to SXDS-27434, a post-starburst galaxy at $z_\mathrm{spec} = 4.013$. The third quiescent galaxy (COS-466654) is not shown for simplicity, though is included in the SED tracks in the leftmost plot.} \label{fig:seds}
\end{figure*} 

% With it's unprecedented combination of sensitivities, resolutions, and a wide field of view, JWST will discover and characterize more $z>3$ quiescent galaxies than ever before. However, at $z\ge3$, rest-frame near-infrared (e.g. $J$-band) spectra are redshifted out of NIRCam's spectral window for galaxies at these epochs. MIRI, with it's smaller field of view, will not fully cover the NIRCam imaging for many of the early JWST legacy surveys currently underway. This is critical because quiescent galaxies at $z>3$ are exceedingly rare \citep[n\,$\sim10^{-5}-10^{-6}$\,Mpc$^{-3}$][]{}, meaning that wide-field surveys with deep, multi-band coverage are critical to detecting statistically significant samples of these objects. Thus, in order to take advantage of this new observational era, new techniques must be developed to rapidly select quiescent galaxies at $z>3$ using the most widely available photometry -- i.e. NIRCam. Recently, \citet{Antwi-Danso2022} proposed a new rest-frame JWST/NIRCam color-color selection technique that will be applicable across a broad range of redshifts, but this method still requires redshift / SED information as a prior to deriving galaxy rest-frame colors. Using zoom in simulations on quiescent galaxies at $z>5$, \citet{Lovell2022} proposed two different observed-frame color-color selection diagnostics with JWST, but both require MIRI photometry and are therefore limited in application. 

This work aims to provide an efficient way to identify $z>3$ quiescent galaxy candidates using only JWST NIRCam photometry. With its unprecedented combination of sensitivity, resolutions, and field of view, JWST will discover and characterize more $z>3$ quiescent galaxies than ever before \cite[and has already demonstrated that promise, e.g. ][]{Carnall2023, Valentino2023}. In order to take advantage of this new observational era, new techniques must be developed to efficiently select quiescent galaxies at $z>3$ using the most widely available and constraining JWST photometry -- i.e. NIRCam imaging. In this \textit{Paper}, we propose an empirical color selection technique for the efficient identification of massive quiescent galaxy candidates at $3 < z < 5$. This method uses the $F277W-F444W$ vs. $F150W-F277W$ color plane (i.e. three bands of photometry -- $F150W, F277W,$ and $F444W$). In addition to its strength in candidate selection, this method is also efficient in that it filters out $>99\%$ of sources, thereby immensely reducing the resources typically dedicated to sifting through catalogs for these rare objects. Furthermore, due to the chosen filters that define this color space, this method is applicable on the majority of ongoing and upcoming JWST wide and/or deep surveys, including COSMOS-Web \citep{Casey2022}, CEERS \citep{Finkelstein2017, Bagley2023}, JADES \citep{Bunker2020}, PANORAMIC \citep{Williams2021}, and PRIMER \citep{Dunlop2021}. 

In Section \ref{sec:sedtemps}, we describe the observed data used to derive and test the $F277W-F444W$ vs. $F150W-F277W$ color space. In Section \ref{sec:method}, we present the selection ``wedge'' and the physical motivation behind the selection criteria. Finally, in Section \ref{sec:appliedtoceers}, we present preliminary results from applying the color selection technique on the Cosmic Evolution Early Release Science (CEERS) Survey. Throughout this work, we adopt a \textit{Planck} cosmology, where $H_0 = 67.7$\,km\,s$^{-1}$\,Mpc$^{-1}$ and $\Omega_\Lambda = 0.692$ \citep{PlanckCollaboration2016}; where relevant, we adopt a Chabrier IMF \citep{chabrier}; all quoted colors / magnitudes are in the AB system. 

% allow the community to swiftly select these objects 
% We propose an \textit{observed-frame} color selection technique to sort through incoming JWST data to quickly isolate the potential massive quiescent galaxy population. While it's not complete / perfect, it will significantly cut down the objects that need filtering. 

\section{SED Templates} \label{sec:sedtemps}

To determine the locus of $3<z<5$ quiescent galaxies, as well as potential contaminant populations, we collate JWST NIRCam colors from a variety of empirically constrained SEDs. The primary set of galaxy SEDs used in this analysis are shown in Figure \ref{fig:seds}. For the purposes of this work, JWST NIRCam colors are interpolated from these SEDs. Future, more expanded tests on the robustness of this selection technique will benefit from results derived from Cycle 1 and Cycle 2 JWST observations. 

\subsection{$z>3$ Quiescent Galaxies}

For known $z = 3-5$ quiescent galaxies, we use the SEDs derived in \citet{Valentino2020} for three spectroscopically-confirmed $z\sim4$ quiescent galaxies, two of which have post-starburst-like SEDs. This is critical to this work since post-starburst galaxies, with young stellar ages and therefore bluer colors, are often excluded in quiescent galaxy studies at $z>3$ as the majority do not fall into the $UVJ$ quiescent region at high-$z$ \citep{Merlin2018, Carnall2020, Stevans2021, Lovell2022}. All three SEDs were modeled with $>10$ photometric measurements covering $\lambda_\mathrm{rest} = 0.4-10$\,$\mu$m, ensuring a well constrained rest-frame UV-to-near-IR SED that traces the shape and strength of the Balmer / 4000\,$\mathrm{\AA}$ break, as well as any UV emission from young stellar populations. We redshift these SEDs from $z=3$ to 6 to chart their evolution through our proposed color space. 

We also include 124 ``robust'' quiescent galaxy candidates at $3<z<5$ identified in \citealt{Gould2023}. These candidates were selected first based on their SED-derived photo-$z$s and stellar masses using photometry from the COSMOS2020 catalog \citep{Weaver2022}, then further required to have $>95\%$ probability of belonging to a quiescent ``group'' as defined by a Gaussian mixture model trained on $2<z<3$ galaxies. The model was trained over several combinations of rest-frame $NUV$, $U$, $V$, and $J$ colors and is more complete in selecting post-starburst galaxies and has less contamination from dusty interlopers when compared to the traditional $UVJ$ method. 

\subsection{Dusty Star Forming Galaxies}

To model the NIRCam colors of dusty, star-forming galaxies (DSFGs, see \citealt*{casey2014dusty}), we use SEDs of ALESS submillimeter galaxies\footnote{\href{http://astronomy.swinburne.edu.au/~ecunha/ecunha/SED_Templates.html}{http://astronomy.swinburne.edu.au/~ecunha/ecunha/\\SED\_Templates.html}} from \citet{dacunha15}. Specifically, we use ALESS SEDs averaged according to varying levels of $V$-band dust attenuation (A$_\mathrm{V}$) to visualize how DSFG NIRCam colors change as a function of attenuation. This is important because the dust-reddened stellar continuum of DSFGs can mimic the red colors of ancient stellar populations in quiescent galaxies and contaminate the quiescent region of $UVJ$ diagrams \citep{Martis2019}. ALESS SEDs were modeled using a rich set of ultraviolet-to-radio photometry and have a median redshift of $z = 2.7\pm0.1$. We use the SED templates corresponding to A$_\mathrm{V} < 1$ (blue / little attenuation), A$_\mathrm{V} \ge 3$ (red / significant attenuation), and the overall average SED with A$_\mathrm{V} \sim 2$. We redshift these SED templates from $z=1$ to 6 to model their color evolution (shown in purple in Figure \ref{fig:seds}). 

% \subsection{Other Potential Contaminants} \label{sec:othercontaminants}

% Studies show that, depending on redshift, strong emission lines can significantly boost the flux captured in wide-band observations and therefore mimic the colors of redder galaxies \citep{Labbe2013, smit16, Schreiber2018}. To test the impact of this effect, we redshift SED templates of galaxies with strong emission lines from $z=1$ to 20 and interpolate their NIRCam colors. We use the latest set of templates from the photo-$z$ fitting code \textsc{eazy} \citep{Brammer2008}, derived from the Flexible Stellar Population Synthesis (FSPS) code \citep{Conroy2010}. We also include an additional set of SED templates \citet{Larson2022}\footnote{\href{https://ceers.github.io/LarsonSEDTemplates}{https://ceers.github.io/LarsonSEDTemplates}} (set 3.5) developed to model strong emission lines emitters and ultra-blue galaxies at high redshift. 

% We find that strong emission line galaxies only enter the 

% Finally, we model active galactic nuclei (AGN) NIRCam colors as evolved from $z=1$ to 10 using SEDs from the SWIRE Template Library.\footnote{\href{http://www.iasf-milano.inaf.it/~polletta/templates/swire\_templates.html}{http://www.iasf-milano.inaf.it/~polletta/templates/swire\_templates.html}} For Type 1 (unobscured) AGN, we used the QSO1 template that combines rest-frame optical to mid-infrared spectra for a sample of spectroscopically confirmed quasars \citep{Hatziminaoglou2005}. For Type 2 (obscured) AGN, we use the IRAS19254-7245 South SED which is a template with a combined starburst and Seyfert 2 AGN component. 

\section{Quiescent Galaxy Selection at $z>3$} \label{sec:method}

In the rest-frame $UVJ$ diagram, the $U-V$ plane spans the Balmer / D$_n$(4000\,$\mathrm{\AA}$) break prominent in galaxies whose light is dominated by aged stellar populations, while the $V-J$ component involves a near-IR detection to break degeneracies between quiescent galaxies and galaxies with spectra reddened by heavy dust obscuration (i.e. DSFGs). However, at $z>3$, the rest-frame $J$-band is redshifted out of NIRCam's spectral window for galaxies at these epochs, and into wavelengths observable by MIRI. Unfortunately, MIRI, with its smaller field of view, will not fully cover the NIRCam imaging for many of the early JWST legacy surveys currently underway (e.g. only the MIRI $F770W$ filter will image $\sim35$\% of the widest Cycle 1 JWST survey, COSMOS-Web,  \citealt{Casey2022}). This is critical because quiescent galaxies at $z>3$ are exceedingly rare \citep[n\,$\sim10^{-5}-10^{-6}$\,Mpc$^{-3}$, ][]{toft14, Valentino2020, Nanayakkara2022, Carnall2023, Valentino2023}, meaning that wide-field surveys with deep, multi-band coverage are critical to detecting statistically significant samples of these objects. Thus, to optimize efficiency and applicability across wide field surveys, we design this selection technique to use NIRCam filters that are available among the majority of recent and upcoming JWST surveys including COSMOS-Web \citep{Casey2022}, CEERS \citep{Finkelstein2017, Bagley2023}, JADES \citep{Bunker2020}, PANORAMIC \citep{Williams2021}, and PRIMER \citep{Dunlop2021}. 

Our first goal is to isolate the Balmer / D$_n$(4000\,$\mathrm{\AA}$) break at $3<z<5$. At these epochs, the Balmer / D$_n$(4000\,$\mathrm{\AA}$) break redshifts to $\lambda_\mathrm{observed} = 1.6-2.4$\,$\mu$m. We wish to define a set of filters that brackets the Balmer break, instead of directly detects it. In other words, the NIRCam $F200W$ band, with a filter throughput that spans $\lambda = 1.7-2.2$\,$\mu$m will not directly measure the Balmer break but instead exhibit significant variance depending on the redshift of the quiescent galaxy. The $F200W$ band is therefore not a reliable tracer at $3<z<5$. Instead, for all quiescent galaxies at $3<z<5$, the Balmer break sits nicely between the $F150W$ and $F277W$ filters (Figure \ref{fig:seds}). This is also advantageous because and the majority of major Cycle 1 JWST surveys have $F150W$ and $F277W$ imaging.

To distinguish between dust-reddened star-forming galaxies and quiescent galaxies, we introduce a third NIRCam band: $F444W$. As shown in Figure \ref{fig:seds}, increasing dust attenuation results in redder $F277W-F444W$ colors, whereas dust-free star-forming galaxies and quiescent galaxies have relatively flat spectral slopes between these two bands. 

In total, we use three NIRCam bands to define this color space: $F150W$, $F277W$, and $F444W$. Using the SED templates described in the previous section, we tested a variety of ``wedges'' in this color space and identified several regions of critical thresholds that have varying degrees of completeness and contamination. In this \textit{Letter}, we share the two versions with the least amount of contamination, while the most complete (but highly contaminated) wedge is listed in the Appendix. In all cases, we require that objects be detected with a signal-to-noise threshold $\ge 3$ in all three NIRCam bands. In order to be considered a quiescent galaxy candidate at $3<z<6$, the most conservative color threshold (aka the ``short wedge'') requires that objects have NIRCam colors that meet the following criteria (all in the AB magnitude system): 

% \begin{gather*} 
%     \mathrm{A.} \indent (F150W - F277W) < 1.5 + 6.25 \times (F277W-F444W) \\
%     \mathrm{and} \\
%     \mathrm{B.} \indent (F150W - F277W) > 1.5 - 0.5 \times (F277W-F444W)\\
%     \mathrm{and} \\
%     \mathrm{C.} \indent (F150W - F277W) > 2.8 \times (F277W-F444W)
% \end{gather*} \label{eqn:1}

\begin{multline}
    \hspace{-0.4cm}\mathrm{A.}\indent(F150W - F277W) < 1.5 + 6.25 \times (F277W-F444W) \\
    \mathrm{and} \\
    \mathrm{B.}\indent (F150W - F277W) > 1.5 - 0.5 \times (F277W-F444W)\\
    \mathrm{and} \\
    \mathrm{C.}\indent (F150W - F277W) > 2.8 \times (F277W-F444W)
\end{multline} \label{eqn:short}

\begin{figure*}[ht!]
\begin{center}
\includegraphics[trim=1cm 1cm 1cm 0cm, width=0.95\textwidth]{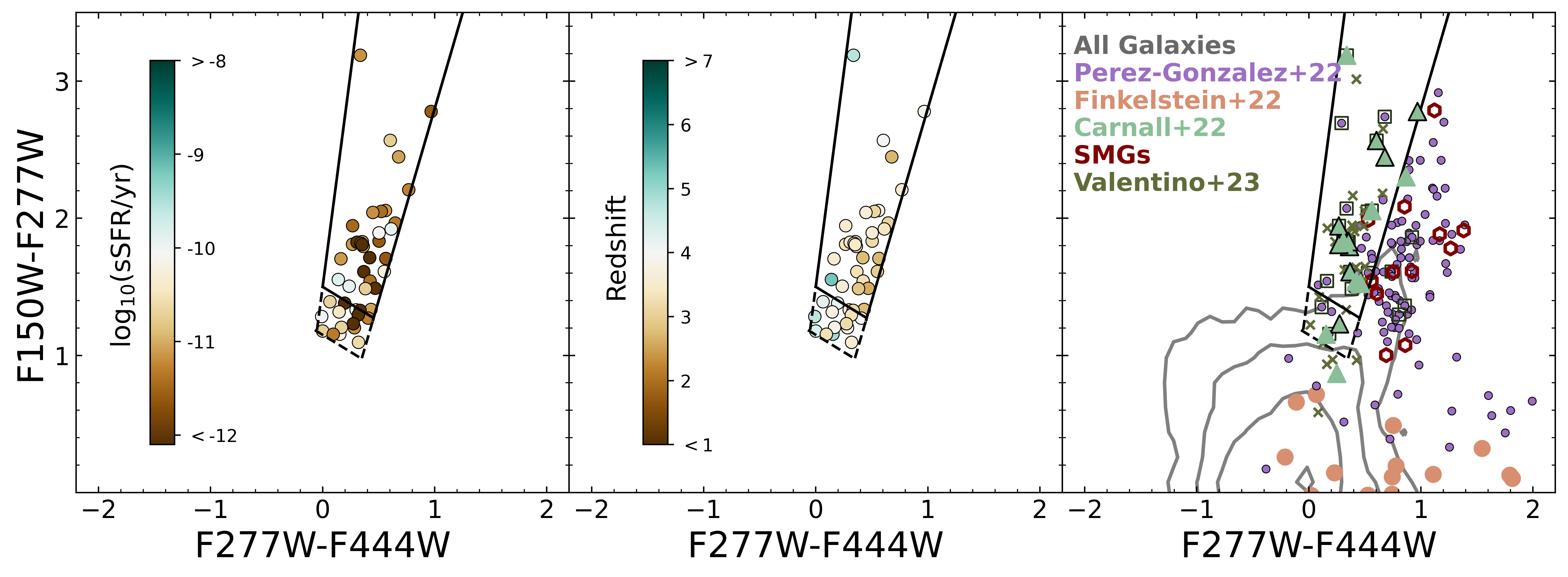}
\end{center}
\caption{$F277W-F444W$ vs. $F150W-F277W$ colors for $2.5 < z < 5$ quiescent galaxy candidates detected in CEERS using the proposed color selection method (including $z_\mathrm{phot}\ge2.5$ and log$_{10}$(sSFR/yr)\,$\lesssim-10$). Interlopers in this color space are further described in Section \ref{sec:contaminants} and shown in Appendix Figure \ref{fig:ceers_interlopers}. \textbf{Left:} The final sample of 44 candidate galaxies colored by sSFRs. \textbf{Middle:} The same sample colored by photometric redshift. \textbf{Right:} All CEERS objects in the $F277W-F444W$ vs. $F150W-F277W$ color space and SNR\,$\ge3$ in all three NIRCam bands (gray contours). Green triangles mark quiescent galaxy candidates from \citet{Carnall2023}, where the most ``robust'' candidates (as defined therein by their sSFR posteriors) are outlined in black. Purple points mark red objects identified by \citet{Perez-Gonzalez2022}, with the quiescent galaxy candidates (also defined by sSFR limits) outlined in black squares. Galaxies with SCUBA-2 and NOEMA detections (i.e. dust obscured) are outlined in red \citep{Zavala2023}. For reference, we also include the ultra-high-$z$ sample of objects characterized in \citet{Finkelstein2023} in pink. } \label{fig:ceers}
\end{figure*}

\noindent The short wedge is represented by solid black lines in Figures \ref{fig:seds} and \ref{fig:ceers}. The short wedge has the most minimal contamination by dust obscured galaxy SED templates, but misses the post-starburst quiescent galaxy SED at $z<4$ as this galaxy's slightly bluer colors push it south of the short wedge's bounds.

To better capture post-starburst galaxies at $z<4$, we introduce a wedge that pushes slightly bluer in both color spaces. This ``long wedge'' requires that Criteria B of the short wedge is changed to:

\begin{multline}
    (F150W - F277W) > \\ 1.15 - 0.5 \times (F277W-F444W)
\end{multline}  \label{eqn:long}

\noindent The long wedge is represented by the additional region of black dashed lines in Figures \ref{fig:seds} and \ref{fig:ceers}. According to the SED templates, the long wedge is potentially more complete in the $z>3$ quiescent galaxy population, but also likely has higher contamination by dusty galaxies and star forming galaxies.  

% \noindent We show this ``wedge'' applied to mock/modeled galaxies (from Section \ref{sec:comparative}) in Figure \ref{fig:color_selection}. We note that the defining bounds of this wedge are somewhat arbitrary. They were chosen in an attempt to capture as many $3<z<5$ quiescent galaxies as possible with little to no contamination from star-forming or $z<2$ quiescent galaxies. However, the wedge's completeness and contamination rates are difficult to robustly quantify as each simulation produces discrepant loci of $3<z<5$ quiescent galaxies, and there is simply not yet enough observational data to test against. In the following Section, we explore this in more detail for the various simulations, and apply this criteria to existing CEERS catalogs as a preliminary proof of concept. 

\section{Applied to CEERS} \label{sec:appliedtoceers}

\subsection{CEERS Data} \label{sec:ceersdata}

In Figure \ref{fig:ceers}, we show our color selection method applied to CEERS NIRCam catalogs, including all ten June and December 2022 pointings. The imaging data and reduction are described in detail in \citet{Bagley2023}. The photometry is performed in two image mode using \textsc{Source Extractor} \citep{sourceextractor} on a (weighted sum) combined F277W and F356W image. Details of the photometry extraction procedure are similar to the method presented in \citet{Finkelstein2023}, with a few updates. Specifically, HST/ACS F606W and F814W mosaics from CANDELS \citep{Koekemoer2011, Grogin2011}, as well as JWST/NIRCam F115W, F150W, and F200W data were convolved to match the PSF of NIRCam F277W imaging. For images with larger PSFs than the F277W imaging (HST/WFC3 F105W, F125W, F140W, and F160W data from CANDELS, as well as NIRCam F356W, F410M, and F444W), we derive a correction favor by convolving the F277W image to the larger PSF, then measuring the flux ratio in the original versus convolved image. This correction factor is applied to fluxes measured in the images with larger PSFs to account for any missing flux in the aperture defined by the F277W image (under the assumption that the morphology does not change significantly). 

Object fluxes were corrected twice more to capture any potential flux missed on larger scales: once by the ratio between the flux measured in small Kron apertures to the default larger \texttt{MAG\_AUTO} Kron aperture, and then an additional $\sim$\,$5-20$\,\% correction to account for missing flux from the wings of the PSF (as determined by source injection simulations). 

The final photometric catalog used in this paper includes the full CEERS suite of imaging data: NIRCam F115W, F150W, F200W, F277W, F356W, F410M, and F444W filters, as well as existing CANDELS HST ACS and WFC3 data in the F606W, F814W, F105W, F125W, F140W, and F160W bands. Photometric redshifts were also calculated for the entire catalog using the method presented in \citet{Finkelstein2023}. 

After filtering for sources with SNR\,$\ge$\,3 in the $F150W$, $F277W$, and $F444W$ bands, we are left with $\sim50$k objects in the CEERS pointings covering $\sim97$\,arcmin$^2$. In the short wedge we retrieve a total of 82 objects, while in the long wedge we retrieve 236 objects (82 of which are the same objects in the short wedge). These selected objects represent $\approx$\,0.2 and 0.5\% of the SNR\,$\ge3$ catalog, respectively, further highlighting the significant reduction in sample size and increase in efficiency when filtering and analyzing large catalogs for high-$z$ quiescent galaxies.

\subsection{SED Fitting}

We use the Code for Investigating GALaxy Emission \citep[\textsc{cigale, }][]{cigale, cigale3, cigale2} to generate galaxy SEDs and photometric redshifts. We assume a \citet{chabrier} initial mass function, \citet{bruz03} stellar population synthesis models, and the \citet{calzetti00} dust attenuation law. The $V$-band attenuation was allowed to vary between $0-5$ magnitudes, and gas and stellar metallicities were fixed to solar ($Z_\odot = 0.02$). We include the option for nebular emission with an ionization parameter between log$_{10}U = [-2, -3]$, and allow the redshift to vary uniformly between $0.8-20$ in steps of $\delta z = 0.1$. We assume a delayed star formation history (``delayed-tau''). For the main stellar population, we fit over a wide range of $e$-folding times ($10-5000$\,Myr). We also allow an optional late burst of star formation, with an $e$-folding time varying between $10-300$\,Myr and a potential fraction of stellar mass created by the late burst to vary between $0-50$\%. 

\begin{deluxetable*}{lccccccccc}[ht]
\centering
\tablecaption{Final sample of 44 quiescent galaxy candidates at $2.5<z<6$ in CEERS.}
 \tablehead{
\colhead{RA} & \colhead{Dec} & \colhead{$m_{150W}$} & \colhead{Redshift} & \colhead{log$_{10}$(M$_\star$/M$_\odot$)} & \colhead{log$_{10}$(sSFR/yr)} & \colhead{Other Ref.$^a$}}
 \startdata
215.01238622 & 53.01419481 & 27.50\,$\pm$\,0.18 & 4.20\,$\pm$\,0.19 & 9.19\,$\pm$\,0.03 & $-10.6$ &      V \\
214.85898311 & 52.89504924 & 25.54\,$\pm$\,0.04 & 4.20\,$\pm$\,0.03 & 10.0\,$\pm$\,0.02 & $-12.0$ &          \\
214.93142047 & 52.93742617 & 25.91\,$\pm$\,0.06 & 4.91\,$\pm$\,0.24 & 9.77\,$\pm$\,0.03 & $-10.1$ &        C \\
214.91938331 & 52.92730122 & 27.03\,$\pm$\,0.08 & 4.68\,$\pm$\,0.18 & 9.31\,$\pm$\,0.02 & $-9.9$ &          \\
214.75157628 & 52.82993207 & 25.24\,$\pm$\,0.02 & 4.00\,$\pm$\,0.02 & 10.2\,$\pm$\,0.02 & $-11.2$ &          \\
214.82580314 & 52.88008725 & 24.19\,$\pm$\,0.02 & 3.59\,$\pm$\,0.01 & 10.2\,$\pm$\,0.05 & $-10.6$ &          \\
214.82613200 & 52.88003104 & 28.43\,$\pm$\,0.16 & 3.78\,$\pm$\,0.37 & 8.79\,$\pm$\,0.04 & $-10.4$ &          \\
214.85799854 & 52.87626021 & 25.55\,$\pm$\,0.12 & 3.87\,$\pm$\,0.13 & 9.84\,$\pm$\,0.02 & $-10.6$ &          \\
214.85621998 & 52.86111432 & 25.62\,$\pm$\,0.04 & 3.61\,$\pm$\,0.04 & 9.88\,$\pm$\,0.02 & $-10.1$ &          \\
214.70744282 & 52.75260243 & 24.25\,$\pm$\,0.02 & 3.10\,$\pm$\,0.02 & 10.4\,$\pm$\,0.02 & $-12.0$ &          \\
214.78830547 & 52.80054129 & 25.42\,$\pm$\,0.03 & 3.89\,$\pm$\,0.02 & 9.89\,$\pm$\,0.02 & $-11.1$ &          \\
214.75193946 & 52.74879797 & 26.43\,$\pm$\,0.05 & 4.32\,$\pm$\,0.11 & 9.43\,$\pm$\,0.02 & $-10.7$ &          \\
214.90955075 & 52.87502532 & 25.18\,$\pm$\,0.02 & 3.30\,$\pm$\,0.02 & 9.99\,$\pm$\,0.02 & $-12.0$ &          \\
214.89561652 & 52.85649304 & 22.93\,$\pm$\,0.01 & 3.19\,$\pm$\,0.02 & 10.8\,$\pm$\,0.02 & $-12.0$ &     C, V \\
215.10403091 & 52.96502357 & 27.40\,$\pm$\,0.13 & 3.29\,$\pm$\,0.18 & 8.91\,$\pm$\,0.04 & $-11.3$ &          \\
214.95787657 & 52.98030101 & 25.13\,$\pm$\,0.02 & 3.53\,$\pm$\,0.15 & 10.4\,$\pm$\,0.04 & $-11.4$ &    PG, C \\
214.98181750 & 52.99123408 & 24.29\,$\pm$\,0.01 & 3.38\,$\pm$\,0.10 & 10.7\,$\pm$\,0.03 & $-11.1$ & PG, C, V \\
215.03905173 & 53.00277846 & 26.49\,$\pm$\,0.07 & 4.00\,$\pm$\,0.56 & 10.5\,$\pm$\,0.07 & $-10.7$ & PG, C, V \\
214.90484984 & 52.93535040 & 24.69\,$\pm$\,0.02 & 3.28\,$\pm$\,0.09 & 10.4\,$\pm$\,0.03 & $-12.7$ & PG, C, V \\
214.86605229 & 52.88425171 & 23.96\,$\pm$\,0.01 & 3.32\,$\pm$\,0.07 & 10.8\,$\pm$\,0.03 & $-11.1$ &     C, V \\
% 214.86604381 & 52.88408282 & 24.86\,$\pm$\,0.02 & 3.73\,$\pm$\,0.18 & 10.7\,$\pm$\,0.04 & $-11.1$ & PG, C, V \\
214.87871537 & 52.88783356 & 26.87\,$\pm$\,0.15 & 3.60\,$\pm$\,0.23 & 9.43\,$\pm$\,0.06 & $-9.9$ &          \\
214.87909817 & 52.88805928 & 25.40\,$\pm$\,0.03 & 3.40\,$\pm$\,0.10 & 10.2\,$\pm$\,0.04 & $-12.0$ & PG, C, V \\
214.76062446 & 52.84531499 & 23.11\,$\pm$\,0.01 & 3.30\,$\pm$\,0.14 & 11.2\,$\pm$\,0.03 & $-11.1$ &     C, V \\
214.83685708 & 52.87344970 & 24.61\,$\pm$\,0.02 & 3.28\,$\pm$\,0.10 & 10.4\,$\pm$\,0.04 & $-11.4$ & PG, C, V \\
214.76722738 & 52.81771171 & 24.94\,$\pm$\,0.03 & 3.53\,$\pm$\,0.18 & 10.5\,$\pm$\,0.04 & $-11.4$ &    PG, V \\
214.85057925 & 52.86601995 & 25.68\,$\pm$\,0.04 & 2.69\,$\pm$\,1.07 & 10.6\,$\pm$\,0.19 & $-11.0$ &    PG, C \\
214.80816482 & 52.83221612 & 27.82\,$\pm$\,0.13 & 4.71\,$\pm$\,0.22 & 10.2\,$\pm$\,0.04 & $-11.1$ & PG, C, V \\
214.76280726 & 52.85128125 & 26.61\,$\pm$\,0.07 & 3.12\,$\pm$\,0.36 & 10.0\,$\pm$\,0.05 & $-11.4$ &       PG \\
214.85390175 & 52.86135518 & 25.26\,$\pm$\,0.04 & 3.86\,$\pm$\,0.29 & 11.3\,$\pm$\,0.06 & $-11.5$ &    PG, C \\
214.79996984 & 52.82209160 & 25.83\,$\pm$\,0.03 & 2.70\,$\pm$\,0.30 & 10.1\,$\pm$\,0.06 & $-11.6$ &       PG \\
214.75817573 & 52.78721770 & 25.64\,$\pm$\,0.04 & 2.99\,$\pm$\,0.13 & 10.2\,$\pm$\,0.06 & $-10.2$ &          \\
214.97856151 & 52.92153875 & 24.89\,$\pm$\,0.03 & 2.61\,$\pm$\,0.14 & 10.4\,$\pm$\,0.03 & $-12.0$ &          \\
214.94173278 & 52.88455850 & 26.15\,$\pm$\,0.05 & 3.06\,$\pm$\,0.39 & 10.3\,$\pm$\,0.06 & $-11.3$ &          \\
214.82773594 & 52.82376795 & 24.22\,$\pm$\,0.02 & 2.85\,$\pm$\,0.19 & 10.5\,$\pm$\,0.05 & $-10.7$ &    PG, V \\
215.06584489 & 52.93295198 & 24.45\,$\pm$\,0.03 & 3.48\,$\pm$\,0.14 & 10.6\,$\pm$\,0.05 & $-12.1$ &          \\
215.11517784 & 52.96071251 & 24.03\,$\pm$\,0.02 & 2.75\,$\pm$\,0.15 & 10.6\,$\pm$\,0.03 & $-11.0$ &          \\
215.02644074 & 52.89377290 & 24.90\,$\pm$\,0.03 & 2.77\,$\pm$\,0.16 & 10.5\,$\pm$\,0.04 & $-12.6$ &          \\
215.04445349 & 52.89882060 & 27.74\,$\pm$\,0.24 & 5.27\,$\pm$\,0.28 & 9.38\,$\pm$\,0.08 & $-9.8$ &          \\
214.98925860 & 52.84716447 & 25.22\,$\pm$\,0.02 & 3.19\,$\pm$\,0.17 & 10.4\,$\pm$\,0.04 & $-11.6$ &          \\
214.89491218 & 52.81715613 & 25.82\,$\pm$\,0.04 & 3.53\,$\pm$\,0.47 & 10.6\,$\pm$\,0.05 & $-11.2$ &          \\
214.93252224 & 52.83243848 & 26.61\,$\pm$\,0.17 & 3.52\,$\pm$\,0.34 & 9.58\,$\pm$\,0.06 & $-11.1$ &          \\
214.97116080 & 52.85489138 & 25.53\,$\pm$\,0.03 & 3.66\,$\pm$\,0.14 & 10.4\,$\pm$\,0.04 & $-11.1$ &          \\
214.89703386 & 52.79221821 & 25.34\,$\pm$\,0.03 & 3.35\,$\pm$\,0.33 & 10.4\,$\pm$\,0.08 & $-9.9$ &          \\
214.77381122 & 52.74001063 & 25.11\,$\pm$\,0.05 & 3.62\,$\pm$\,0.51 & 10.5\,$\pm$\,0.09 & $-10.0$ &          \\
 \enddata  \tablenotetext{a}{PG corresponds to sources captured in \citet{Perez-Gonzalez2022}, C corresponds to sources listed in \citet{Carnall2023}, and V corresponds to sources listed in \citet{Valentino2023}.}
\end{deluxetable*} \label{table:candidates}
% 214.751547 +52.830072
\subsection{Quiescent Galaxies in CEERS} \label{sec:qgsinceers}

When using only photometry to identify high-$z$ quiescent galaxies, issues with sample completeness and contamination are perhaps mitigated best by defining thresholds in galaxy specific star formation rates (sSFRs\,$=$\,SFR/M$_\star$). The majority of star-forming galaxies lie on a well-established SFR-M$_\star$ relationship \citep[e.g.][]{Noeske2007a, speagle}, with quiescent galaxies falling $1-2$\,dex below this plane \citep[e.g.][]{Straatman2014, Pacifici2016}. Working with a sSFR threshold is likely more advantageous at $z>3$ as this method provides more flexibility in capturing galaxies that are either in the process of quenching (``green valley'' objects), or recently quenched objects with leftover UV emission from the last generation of young stars (``post-starbursts'').

We select high-$z$ quiescent galaxy candidates as objects with $z_\mathrm{phot}\ge2.5$ and log$_{10}$(sSFR/yr)\,$\lesssim-10$, and present these candidates in Table \ref{table:candidates}. In the short wedge, this yields a total of 30 candidates, while the long wedge adds another 15 candidates, making a total of 45 high-$z$ quiescent galaxy candidates. As mentioned later in Section \ref{sec:contaminants}, we remove a single source that is coincident within 1'' of a submillimeter source and is therefore likely a heavily-obscured interloper. The remaining 44 sources span $z=2.6-5.3$ with a median redshift of $z\sim3.5$. They are relatively massive with a median stellar mass of M$_\star \approx 3 \times 10^{10}$\,M$_\odot$, and quiescent with a median log$_{10}$(sSFR/yr)\,$\approx -11.2$. We also note that all but one of our candidates have $\ge$3$\sigma$ detections at $m_{F150W} < 28$, while the CEERS Survey reaches 5$\sigma$ depths of $m_{F150W} \sim 29$ \citep{Bagley2023}; therefore, this color selection method can be safely and successfully applied to shallower JWST Surveys (e.g. COSMOS-Web with 5$\sigma$ depths up to $m_{F150W} \approx 28$, \citealt{Casey2022}). 

As shown in Figure \ref{fig:ceers}, nearly all (14/15) of the CEERS quiescent galaxy candidates presented in \citet{Carnall2023} are recovered in our analysis (both by nature of their loci on the color diagram, but then again by our SED fitting procedure). Out of the 25 quiescent galaxy candidates in \citet{Perez-Gonzalez2022}, 22 meet the SNR thresholds, and only 18/22 fall in the long wedge. Our SED results recover 11 of the 18 candidates as quiescent galaxies $z \ge 2.5$. We find that 3 galaxies designated as star forming in P\'{e}rez-Gonz\'{a}lez et al. are also recovered as quiescent in this work, though our analysis prefers lower redshift solutions for two of these objects ($\delta z \sim 1$) than those presented in P\'{e}rez-Gonz\'{a}lez et al., which is likely driving the discrepant star forming properties. The remaining quiescent candidates in P\'{e}rez-Gonz\'{a}lez et al. are fit as coeval star forming (or even bursting) galaxies in our SED fitting process with log$_{10}$(sSFR/yr)\,$\sim - 9$ to $-8$ and A$_\mathrm{V} \sim 2-4$. For the majority of objects that overlap as quiescent galaxy candidates in Carnall et al. and/or P\'{e}rez-Gonz\'{a}lez et al., our analysis produces photometric redshifts within $\Delta z \approx 0.5$. 

We also compare against the 24 CEERS quiescent galaxy candidates from \citet[][priv. communication]{Valentino2023} and find that 19 of these candidates fall into the long wedge space. The five sources that fall outside the wedge have lower stellar masses than those within the wedge (with $<\mathrm{log_{10} M}_\star> = 9.45$\,$M_\sun$ and 10.74\,$M_\sun$, respectively) and were reported to show features signifying more recent quenching. Of the objects that fall into the wedge, we recover 14 quiescent galaxy candidates at $z\gtrsim3$. The remaining five objects are fit as coeval starburst-like galaxies with log$_{10}$(sSFR/yr)\,$= - 9$ to $-8$ and A$_\mathrm{V} \sim 2-3$, further highlighting the difficulty in dividing these two galaxy populations with photometry alone. 

Perhaps more notable than the recovery of previously identified high-$z$ quiescent galaxy candidates is the discovery of many new ones. When we reduce our candidate pool to objects discovered in the same CEERS pointings as those presented in \cite{Perez-Gonzalez2022}, \cite{Carnall2023}, and \cite{Valentino2023} -- specifically pointings 1, 2, 3 and 6 -- we identify an additional $13-14$ new $z>3$ quiescent galaxy candidates. This is a near doubling of the candidate population. Moreover, the average specific star formation rate for the newly discovered population is log$_{10}$(sSFR/yr)\,$=-10.4 \pm 0.4$, but for the previously discovered population it is log$_{10}$(sSFR/yr)\,$=-11.1 \pm 0.4$. These population characteristics suggest that this color selection technique is successful in capturing young, post-starburst quiescent galaxies as well as more evolved galaxies with little to no young stellar populations. Such a population yield is critical as most quiescent galaxies at these epochs are expected to be post-starbursts \cite[with log$_{10}$(sSFR/yr)\,$=-10$ to $-11$, ][]{D'Eugenio2020}.

More details on the properties of the quiescent galaxy candidate sample in this work and across other public JWST Cycle 1 surveys will be presented in a forthcoming paper (Long et al. in prep).

\begin{figure}[ht!]
\includegraphics[trim=0.5cm 0.5cm 0.5cm 0cm, width=0.45\textwidth]{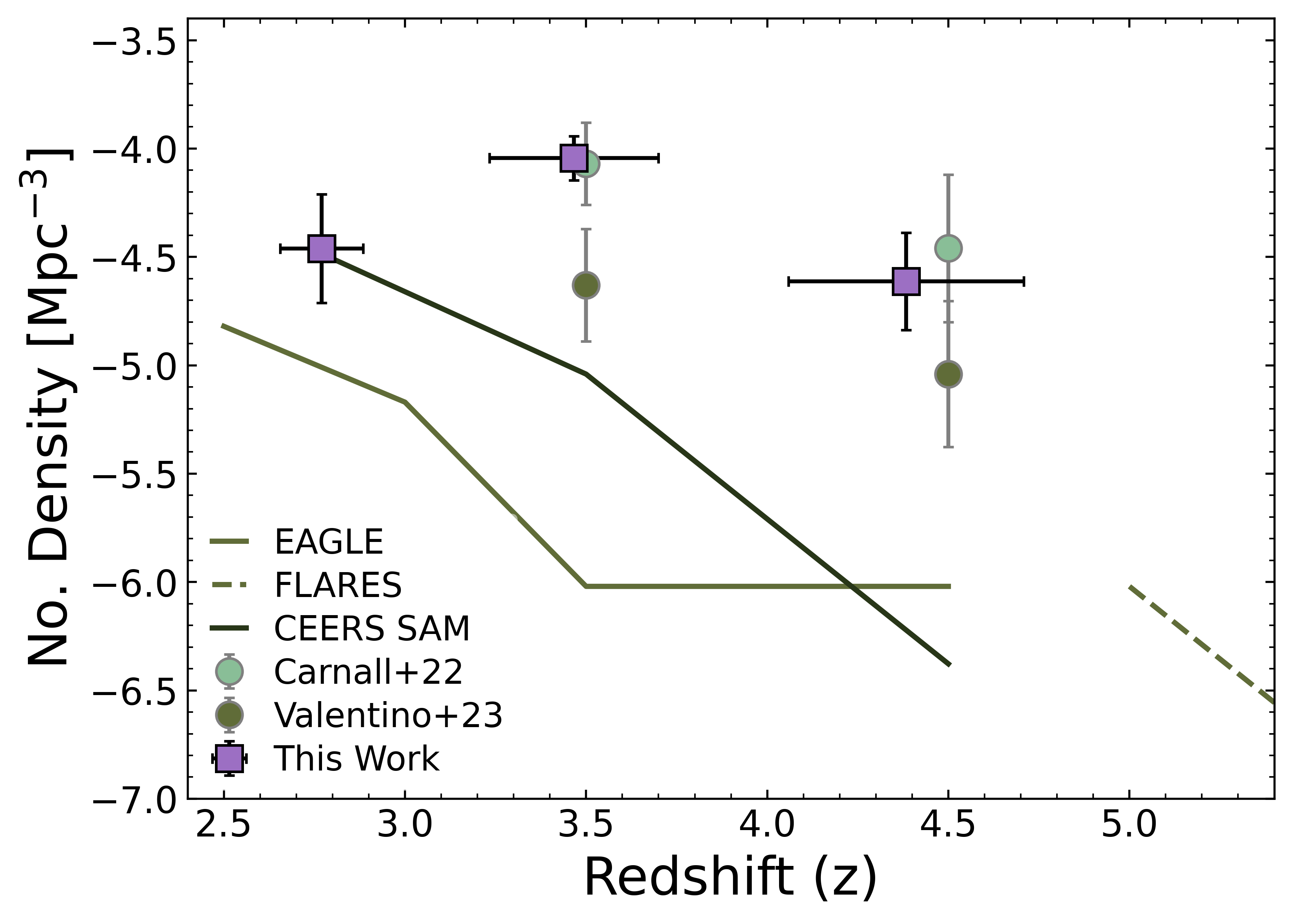}
\caption{Number density evolution derived using the sample of quiescent galaxies identified in this work (purple squares, also listed in Table \ref{table:nodens}). The other two sets of points are reported number densities of quiescent galaxies derived also using JWST data \citep{Carnall2023, Valentino2023}; the discrepancy between our estimates and \citet{Valentino2023} are likely due to differences in stellar mass thresholds (discussed further in Section \ref{sec:qgsinceers}). We also overlay predictions from several cosmological simulations described in Section \ref{sec:simulations} and shown in Figure \ref{fig:sim_color_selection}.} \label{fig:nodensity}
\end{figure}

In Figure \ref{fig:nodensity}, we derive number densities across redshift ranges of $z = [2.5, 3), [3, 4)$, and $[4,5)$. To derive conservative uncertainties on these estimates, we ran 10$^3$ Monte Carlo simulations that sampled each object's redshift from a uniform prior defined by their photo-$z$ estimate $\pm$ their respective 1$\sigma$ uncertainties. We calculate the number density in each redshift bin for each realization, and then report the median number densities for this work. Uncertainties were derived from the inner 68\% confidence interval of the MCMC-computed values plus Poisson noise. These estimates are presented in Table \ref{table:nodens}. 

We compare our number density estimates to other literature that includes JWST data in their selection of massive quiescent galaxies -- namely, \citet{Carnall2023} and \citet{Valentino2023}. Our estimates are in excellent agreement with those derived in \citet{Carnall2023}, despite different selection techniques over similar fields / data sets, but $\sim3-5\times$ greater than estimates presented in \citet{Valentino2023}. The former is unsurprising as we successfully recover 14/15 of the quiescent galaxy candidates in Carnall et al., and our sample is expanded over 2.5$\times$ the area analyzed therein. The discrepancy with Valentino et al. is likely due to a stellar mass threshold: if we reduce our sample to candidates that meet the criteria for stellar mass therein (log$_{10}$\,M$_\star \ge 10.6$), then we achieve near identical number density estimates. Therefore, the population densities derived via our proposed color selection technique (the long wedge, specifically), when combined with SED fitting, appears consistent with other high-$z$ quiescent galaxy selection techniques. In a forthcoming paper, we will explore how these number densities vary as a function of specific star formation rates, stellar mass, and more with a larger sample of quiescent galaxy candidates selected using this method.

\subsection{Contaminants} \label{sec:contaminants}

% AGN:
% We find that the type 1 (unobscured) AGN do not enter the quiescent region of our proposed color space. Type 2 (obscured) AGN occupy a similar color space as DSFGs, which is unsurprising as type 2 AGN tend to live in heavily dust-obscured systems. Therefore, AGN as an individual population are unlikely a primary contaminant in this color space, and will likely only enter the wedge when embedded within and therefore may represent a contaminant sub-population since its likely most if not all of heavily obscured AGN are in heavily obscured galaxies. 

% Type 1: QSO1
% The type of AGN that contaminate the wedge are the ones that live in heavily dust obscured systems, which are already identified as a contaminant population. Specifically, obscured AGN in DSFGs.

In the long (short) wedge, there are 191 (52) objects that do not meet the criteria for quiescent galaxy candidates at $z>2.5$. We describe below some of the properties of these contaminant galaxies, and show in the Appendix some of their SED-derived properties. At a high level, there is unfortunately no clear threshold in flux/magnitude across HST and JWST bands that easily separates the contaminants from the high-$z$ quiescent galaxy candidates. Future spectroscopically confirmed samples will certainly help identify any potential additional thresholds that could be applied to further reduce the contaminant population.

The majority ($\gtrsim$\,80\%) of the contaminant objects are fit as emission line / starburst-like galaxies at $z\lesssim6$. Redshifted nebular emission lines (such as H$\alpha$ and H$\beta$) can contribute significant flux to wide band passes, thereby making a galaxy appear redder than they truly are \citep{Labbe2013, smit16, Schreiber2018, Antwi-Danso2022, McKinney2023}. For visual demonstration of contamination in the color space, we redshift SED templates of galaxies with strong emission lines from $z=1$ to 20 using the latest set of templates from the photo-$z$ fitting code \textsc{eazy} \citep[][derived from the Flexible Stellar Population Synthesis (FSPS) code in \citealt{Conroy2010}]{Brammer2008}. We include an additional set of SED templates \citet[][set 3.5]{Larson2022}\footnote{\href{https://ceers.github.io/LarsonSEDTemplates}{https://ceers.github.io/LarsonSEDTemplates}} developed to model strong emission lines emitters and ultra-blue galaxies at high-$z$. We show these interpolated colors in Figure \ref{fig:seds}. We find that emission line galaxies enter the long wedge color space at $z \sim 4-5$ and $z\sim14$. The latter ultra-high redshift population can be removed using a magnitude threshold in the $F115W$-band: nearly all (25/26) of the $z>8$ candidate galaxies presented in \citet{Finkelstein2023} have m$_{F115W} < 28$ (AB mag), while all but 1 of our candidate quiescent galaxies have m$_{F115W} > 28$ (AB mag). 

We performed a similar analysis to assess for potential contamination from active galactic nuclei (AGN). We evolved AGN SED templates using SEDs from the SWIRE Template Library\footnote{\href{http://www.iasf-milano.inaf.it/~polletta/templates/swire\_templates.html}{http://www.iasf-milano.inaf.it/~polletta/templates/swire\_templates.html}} from $z=1$ to 10. For Type 1 (unobscured) AGN, we used the QSO1 template that combines rest-frame optical to mid-infrared spectra for a sample of spectroscopically confirmed quasars \citep{Hatziminaoglou2005}. For Type 2 (obscured) AGN, we use the IRAS19254-7245 South SED which is a template with a combined starburst and Seyfert 2 AGN component. We find that the type 1 (unobscured) AGN do not enter the quiescent region of our proposed color space as they are too blue. Type 2 (obscured) AGN occupy a similar color space as DSFGs, which is unsurprising as type 2 AGN tend to live in dust-obscured systems. Thus, AGN as an individual population are unlikely a primary contaminant in this color space, but instead may represent a contaminant \textit{sub}-population. 

As shown in Appendix Figure \ref{fig:ceers_interlopers} and predicted in Figure \ref{fig:seds}, heavily dust obscured galaxies (with A$_V \ge 1$) lie primarily outside of the wedge due to their redder $F277W-F444W$ colors. This is also illustrated by the colors of the 14 submillimeter sources identified via SCUBA-2 observations in the field \citep[seen in our Figure \ref{fig:ceers}, ][]{Zavala2017, Zavala2018ceers, Zavala2022}. Still, some contamination is expected since dust-reddened spectra can mimic the red colors of aged stellar populations. In the long (short) wedge, there are roughly 40 (22) objects with significant attenuation (A$_V \ge 1$) such that their potentially dust-reddened spectra pushes these objects into this color space. Nearly all ($\gtrsim$\,90\%) of these dusty objects are predicted to sit at $z<3$, though a handful have photo-$z$ solutions at $z\sim4-5$. Only one of the 14 dusty, star-forming galaxies identified in \citet[][priv. communication]{Zavala2023} is captured in our wedge, demonstrating that the majority of potential dust-obscured contaminants may have low IR luminosities (L$_\mathrm{IR} \lesssim 10^{12}$\,L$_\odot$). This same object is deemed a quiescent galaxy by our SED photo-$z$ fitting procedure, but we remove it from the final reported sample. Unfortunately, a more explicit quantification of this dust-obscured contamination rate requires additional data (e.g. ALMA or JWST) to fully confirm the nature of both the contaminants and the candidates. 

Finally, we also find two galaxies with photo-$z$ solutions at $z\sim11$ within the wedge. Both of these objects have marginal or no detections bluewards of the $F150W$ band. One is also detected in P\'{e}rez-Gonz\'{a}lez et al. as a quiescent galaxy candidate at $z\sim4$. Based on the redshifted SED templates discussed above, star forming galaxies only enter this color space in specific redshift windows of $z\sim3-5$ and $z\gtrsim14$ due to emission lines for the former and significant Lyman breaks for the latter. This includes heavily dust-reddened spectra. Furthermore, the photo-$z$ uncertainty on these objects is large ($\Delta z \sim 2$). Thus, we urge the reader to interpret the validity of these two contaminants with caution. 

\begin{figure*}[ht!]
\begin{center}
\includegraphics[trim=0cm 1cm 0cm 0cm, width=1\textwidth]{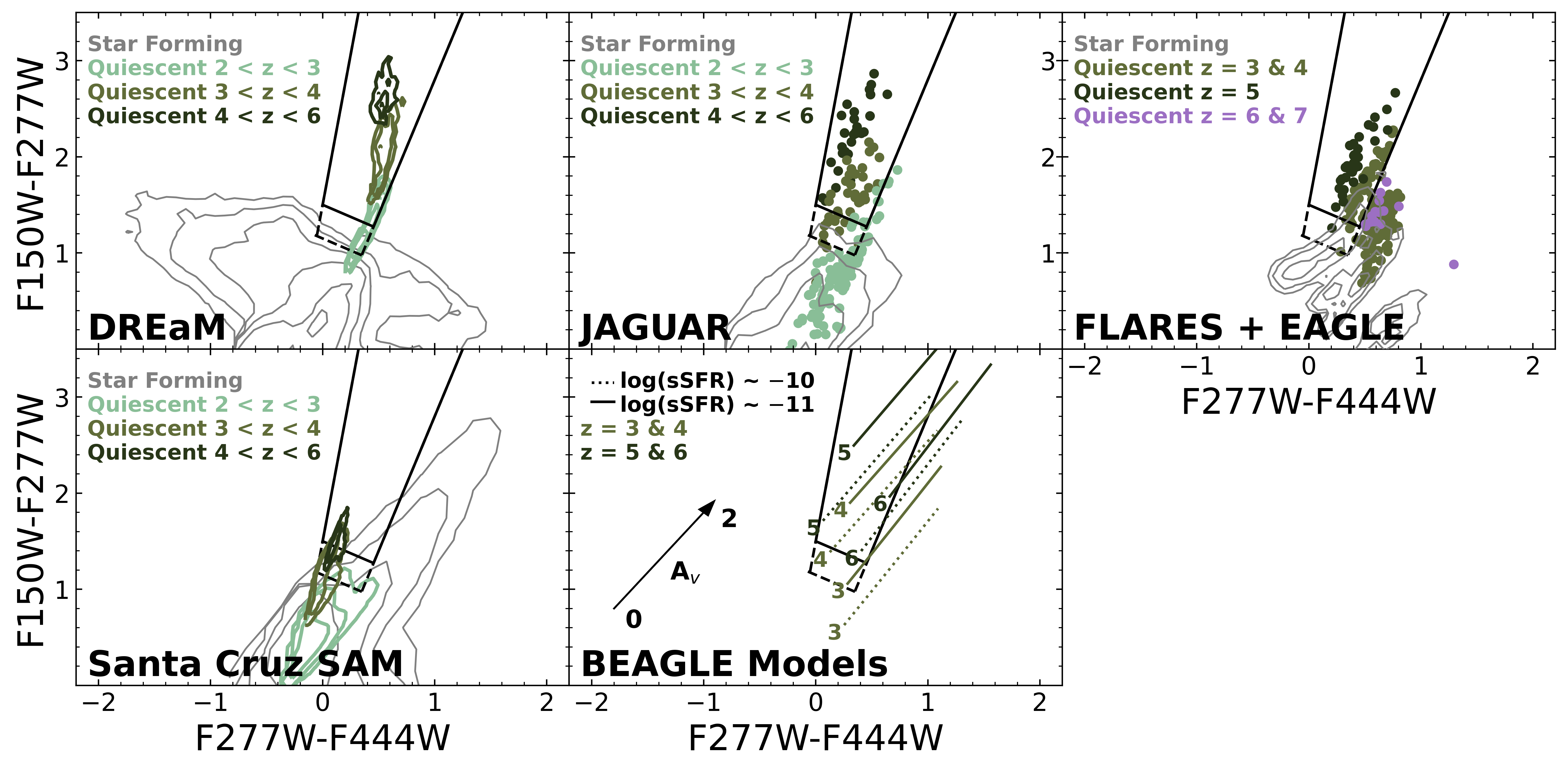}
\end{center}
\caption{$F277W-F444W$ vs. $F150W-F277W$ colors for quiescent galaxies across several simulations. Black solid and dashed lines are the same ``short'' and ``long'' wedges denoted in previous figures. \textbf{Top:} Mock galaxy colors from the DREaM semi-analytical model \citep[left,][]{Drakos2022}, the JAGUAR semi-analytical model \citep[middle,][]{Williams2018}, and the FLARES\,$+$\,EAGLE \citep[right,][]{Vijayan2021, Lovell2021a, Lovell2022} hydrodynamical simulation with quiescent populations highlighted in various shades of green and purple to mark their redshift ranges, with the remaining galaxies shown in grey contours. \textbf{Bottom:} Mock galaxy colors from the Santa Cruz semi-analytical model \citep[left,][]{Yung2022} with a similar key as the figures in the top panel. On the right are mock galaxy colors generated by \textsc{BEAGLE} \citep{Chevallard2016}. Details on how these colors are generated are presented in Section \ref{sec:simulations}. Briefly, we show the color evolution of both aged (solid lines) and post-starburst (dotted lines) galaxies with varying levels of attenuation (A$_\mathrm{V} = 0-2$) at $z=3$, 4, 5, and 6. } \label{fig:sim_color_selection}
\end{figure*}

\begin{deluxetable}{ccc}[ht!]
\centering
\tablecaption{Number density of quiescent galaxies identified in this work.}
 \tablehead{
\colhead{Redshift Range} & \colhead{Number} & \colhead{$n$/Mpc$^{-3}$} }
 \startdata
 $2.5 < z < 3$ & 7 & 2.88\,$^{+2.24}_{-1.96} \times 10^{-5}$\\
 $3 < z < 4$ & 29 & 1.09\,$^{+0.28}_{-0.23} \times 10^{-5}$\\
 $4 < z < 5$ & 8 & 4.07\,$^{+2.69}_{-1.62} \times 10^{-5}$\\
 \enddata 
\end{deluxetable} \label{table:nodens} \vspace{-2cm}

\section{Applied to Simulations} \label{sec:simulations}

We apply the proposed color selection technique on a variety of cosmological simulations with mock JWST photometry, as shown in Figure \ref{fig:sim_color_selection}. Specifically, we apply this color selection criteria to the Deep Realistic Extragalactic Model (DREaM) semi-analytic model \citep{Drakos2022}, Santa Cruz semi-analytic model (SAM) \citep{Somerville2015b, Yung2019a, Somerville2021, Yung2022}, the JAdes extraGalactic Ultradeep Artificial Realizations phenomenological model \citep[JAGUAR, ][]{Williams2018}, and the \textsc{eagle} and \textsc{flares} hydrodynamical simulation \citep{Vijayan2021, Lovell2021a, Lovell2022}. We refer the readers to the above references for details on these models and their assumptions. For visual comparison, we also include mock quiescent galaxy SED models generated via the BEAGLE tool \citep{Chevallard2016} using the \citet{bruz03} templates; these mock SEDs are at solar metallicity, with one model representing an older (1\,Gyr) stellar population, and the other representing a younger (500\,Gyr) stellar population, both generated using a delayed-$\tau$ SFH where $\tau = 100$ or 60\,Myr, respectively. We also apply varying levels of attenuation (A$_\mathrm{V} = 0-2$) to examine the effects of dust on quiescent galaxy colors. 

Note that we apply the same definitions and requirements for high-$z$ quiescent galaxy candidacy: objects must sit at $z \ge2.5$ and exhibit specific star formation rates of log$_{10}$(sSFR/yr)\,$\lesssim-10$. 

In general, $z\ge3$ quiescent galaxy populations across the simulations are well captured by the long wedge. The empirical and semi-analytic models (DREaM, JAGUAR, and Santa Cruz SAM) have simulated galaxy spectra that are fairly red due to perhaps their dust prescriptions or to the fact that they assign the SEDs of heavily evolved galaxies to their quiescent samples \citep[e.g.,][]{Somerville2021}; this is likely why the majority of high-$z$ quiescent galaxies in these simulations fall neatly into the wedge space. Future work on how well this wedge captures young, post-starburst quiescent galaxies will be needed. In the \textsc{eagle} and \textsc{flares} simulations, the main population missed by the wedge is $z\sim3$ quiescent objects, which is in line with the BEAGLE model predictions (whether post-starburst or strongly evolved). Considering that the dominant contaminant population in all of these simulations is heavily dust obscured galaxies (with A$_\mathrm{V}\gtrsim1$), and that each simulation has a unique prescription for dust production, attenuation, and/or extinction, it is clear that a better understanding of the prevalence of dust in both star-forming and quiescent galaxies is necessary. Such an analysis is beyond the scope of this paper and saved for future work.

\section{Discussion \& Summary}
As mentioned in Section \ref{sec:method}, our proposed color selection technique is not entirely distinct conceptually from several of the new rest-frame color selection techniques proposed in the literature. For example, when redshifted to $3<z<5$ the rest-frame synthetic ($ugi$)$_s$ filters explored in \citet{Antwi-Danso2022} constrain generally similar parts of the spectrum as the NIRCam $F150W$, $F277W$, and $F444W$ bands proposed in this work \citep[see also e.g. ][]{Liu2018}. The $NUVU-VJ$ color diagram presented in Gould et al. is also similar at these redshifts, however the rest-frame $J$-band at $z>3$ presents a data challenge where MIRI observations are required to truly constrain this part of the spectrum. MIRI, with its smaller field of view and limited sensitivities, will not fully cover the NIRCam imaging for several of the early JWST legacy surveys currently underway. This is critical as quiescent galaxies at $z>3$ are exceedingly rare \citep[n\,$\sim10^{-5}-10^{-6}$\,Mpc$^{-3}$][]{Girelli2019, Santini2019, Shahidi2020, Valentino2020, Long2022, Carnall2023}, meaning that wide-field surveys with uniform multi-band coverage are necessary to detect statistically significant samples of these objects. \citet{Lovell2022} also proposed observed-frame color selection diagnostics with JWST, however, both require MIRI photometry and are therefore currently limited to smaller data sets with these specific bands.

The strength of our proposed technique lies in its empirical nature, enabled by the filters on JWST's NIRCam instrument alone. With this technique, the initial labor in identifying these objects is significantly reduced: one does not need to generate photometric redshifts and/or SED catalogs for tens of thousands of objects as a prior step in the search for quiescent galaxies at $3<z<5$. Instead, it is clear from Figure \ref{fig:ceers} and \ref{fig:sim_color_selection} that red objects (whether quiescent or dust obscured) occupy a specific region in the $F277W-F444W$ vs. $F150W-F277W$ color space, and that this can be exploited to identify a much smaller pool of candidates to pull from (i.e. $<1\%$ of all detected objects). This will undoubtedly increase our efficiency in discovering quiescent galaxies at high-$z$.

Furthermore, this technique also shows potential for capturing young, massive galaxies in the throes of quenching -- i.e. post-starbursts. Quiescent galaxies in the first 2\,Gyr often have small, but significant amounts of UV light from their recent starburst episode \cite[e.g. ][]{D'Eugenio2020, Marsan2022}. These slightly bluer colors are likely why many $z>3$ quiescent galaxies can be missed in searches that use classic $z<2$ quiescent galaxy selection techniques, as they are tuned to find red galaxies with no young stellar populations. As shown in Section \ref{sec:appliedtoceers}, the average specific star formation rate of newly discovered quiescent galaxy candidates in this work is roughly 5$\times$ (0.7\,dex) lower than those that have already been identified using the low-$z$ quiescent galaxy color selection techniques. Therefore, this color plane demonstrates potential as a more complete selection method as it captures both mature and recently quenched massive galaxies at $z>3$. However, larger samples are needed to explore and quantify its completeness at a statistical level.

We use empirically-constrained galaxy SEDs to derive an observed-frame color selection technique for massive quiescent galaxies at $3\gtrsim z \lesssim5$ using JWST NIRCam imaging. Our $F277W-F444W$ vs. $F150W-F277W$ color selection method is similar in concept to well known color selection techniques in the literature (e.g. $UVJ$) but is more efficient and advantageous as it does not require any prerequisite photo-$z$ and/or SED fitting, and also captures more young post-starburst galaxies than techniques tuned to the low redshift Universe. We demonstrate the efficacy of this method by applying this technique to JWST imaging in the CEERS field: we identify 44 quiescent galaxy candidates at $2.5<z<6$. We recover nearly all quiescent galaxies at this epoch previously identified in the literature, and also discover 26 new candidates, the majority of which are likely post-starbursts. Similar to other color selection techniques, this technique also suffers from contamination from heavily dust-obscured sources, though the quantification of this false-positive rate requires additional data (e.g. ALMA or JWST) to fully confirm the nature of both the contaminants and the candidates. Future, more refined versions of this technique will be developed upon the availability of additional wide-field observations with multiwavelength data (e.g. COSMOS-Web, \citealt{Casey2022}).

\section{Acknowledgements}

This work is dedicated to little Black girls whose their hearts are set alight by scientific inquiry and exploration. May you live lives full of curiosity, and may the world never dull that light within you.

We honor the invaluable labor of the maintenance and clerical staff at our  institutions, whose contributions make our scientific discoveries a reality. This work was developed and written in central Texas on the lands of the Tonkawa, Comanche, and Apache people. 

ASL would like to thank Charlie and Patrick Long for the love, support, and precious moments baking in the sun together. ASL would also like to thank Rachel Nere for her incisive curiosity surrounding JWST and galaxy evolution during the Summer 2022 TAURUS program. ASL acknowledges support for this work provided by NASA through the NASA Hubble Fellowship Program grant \#HST-HF2-51511.001-A, awarded by the Space Telescope Science Institute, which is operated by the Association of Universities for Research in Astronomy, Inc., for NASA, under contract NAS5-26555.

This work is based [in part] on observations made with the NASA/ESA/CSA JWST. The data were obtained from the Mikulski Archive for Space Telescopes at the Space Telescope Science Institute, which is operated by the Association of Universities for Research in Astronomy, Inc., under NASA contract NAS 5-03127 for JWST. These observations are associated with program JWST \#1345.

Some of the data presented in this paper are available on the Mikulski Archive for Space Telescopes (MAST) at the Space Telescope Science Institute. The specific observations can be accessed via doi:10.17909/qhb4-fy92 and doi:10.17909/T94S3X.

This work made use of Astropy:\footnote{http://www.astropy.org} a community-developed core Python package and an ecosystem of tools and resources for astronomy \citep{astropy:2013, astropy:2018, astropy:2022}. 
\bibliography{references}{}

\begin{thebibliography}{}
\expandafter\ifx\csname natexlab\endcsname\relax\def\natexlab#1{#1}\fi
\providecommand{\url}[1]{\href{#1}{#1}}
\providecommand{\dodoi}[1]{doi:~\href{http://doi.org/#1}{\nolinkurl{#1}}}
\providecommand{\doeprint}[1]{\href{http://ascl.net/#1}{\nolinkurl{http://ascl.net/#1}}}
\providecommand{\doarXiv}[1]{\href{https://arxiv.org/abs/#1}{\nolinkurl{https://arxiv.org/abs/#1}}}

\bibitem[{{Antwi-Danso} {et~al.}(2022){Antwi-Danso}, {Papovich}, {Leja},
  {Marchesini}, {Marsan}, {Martis}, {Labb{\'e}}, {Muzzin}, {Glazebrook},
  {Straatman}, \& {Tran}}]{Antwi-Danso2022}
{Antwi-Danso}, J., {Papovich}, C., {Leja}, J., {et~al.} 2022, arXiv e-prints,
  arXiv:2207.07170.
\newblock \doarXiv{2207.07170}

\bibitem[{{Astropy Collaboration} {et~al.}(2013){Astropy Collaboration},
  {Robitaille}, {Tollerud}, {Greenfield}, {Droettboom}, {Bray}, {Aldcroft},
  {Davis}, {Ginsburg}, {Price-Whelan}, {Kerzendorf}, {Conley}, {Crighton},
  {Barbary}, {Muna}, {Ferguson}, {Grollier}, {Parikh}, {Nair}, {Unther},
  {Deil}, {Woillez}, {Conseil}, {Kramer}, {Turner}, {Singer}, {Fox}, {Weaver},
  {Zabalza}, {Edwards}, {Azalee Bostroem}, {Burke}, {Casey}, {Crawford},
  {Dencheva}, {Ely}, {Jenness}, {Labrie}, {Lim}, {Pierfederici}, {Pontzen},
  {Ptak}, {Refsdal}, {Servillat}, \& {Streicher}}]{astropy:2013}
{Astropy Collaboration}, {Robitaille}, T.~P., {Tollerud}, E.~J., {et~al.} 2013,
  \aap, 558, A33, \dodoi{10.1051/0004-6361/201322068}

\bibitem[{{Astropy Collaboration} {et~al.}(2018){Astropy Collaboration},
  {Price-Whelan}, {Sip{\H o}cz}, {G{\"u}nther}, {Lim}, {Crawford}, {Conseil},
  {Shupe}, {Craig}, {Dencheva}, {Ginsburg}, {VanderPlas}, {Bradley},
  {P{\'e}rez-Su{\'a}rez}, {de Val-Borro}, {Aldcroft}, {Cruz}, {Robitaille},
  {Tollerud}, {Ardelean}, {Babej}, {Bach}, {Bachetti}, {Bakanov}, {Bamford},
  {Barentsen}, {Barmby}, {Baumbach}, {Berry}, {Biscani}, {Boquien}, {Bostroem},
  {Bouma}, {Brammer}, {Bray}, {Breytenbach}, {Buddelmeijer}, {Burke},
  {Calderone}, {Cano Rodr{\'{\i}}guez}, {Cara}, {Cardoso}, {Cheedella},
  {Copin}, {Corrales}, {Crichton}, {D'Avella}, {Deil}, {Depagne}, {Dietrich},
  {Donath}, {Droettboom}, {Earl}, {Erben}, {Fabbro}, {Ferreira}, {Finethy},
  {Fox}, {Garrison}, {Gibbons}, {Goldstein}, {Gommers}, {Greco}, {Greenfield},
  {Groener}, {Grollier}, {Hagen}, {Hirst}, {Homeier}, {Horton}, {Hosseinzadeh},
  {Hu}, {Hunkeler}, {Ivezi{\'c}}, {Jain}, {Jenness}, {Kanarek}, {Kendrew},
  {Kern}, {Kerzendorf}, {Khvalko}, {King}, {Kirkby}, {Kulkarni}, {Kumar},
  {Lee}, {Lenz}, {Littlefair}, {Ma}, {Macleod}, {Mastropietro}, {McCully},
  {Montagnac}, {Morris}, {Mueller}, {Mumford}, {Muna}, {Murphy}, {Nelson},
  {Nguyen}, {Ninan}, {N{\"o}the}, {Ogaz}, {Oh}, {Parejko}, {Parley}, {Pascual},
  {Patil}, {Patil}, {Plunkett}, {Prochaska}, {Rastogi}, {Reddy Janga},
  {Sabater}, {Sakurikar}, {Seifert}, {Sherbert}, {Sherwood-Taylor}, {Shih},
  {Sick}, {Silbiger}, {Singanamalla}, {Singer}, {Sladen}, {Sooley},
  {Sornarajah}, {Streicher}, {Teuben}, {Thomas}, {Tremblay}, {Turner},
  {Terr{\'o}n}, {van Kerkwijk}, {de la Vega}, {Watkins}, {Weaver}, {Whitmore},
  {Woillez}, {Zabalza}, \& {Astropy Contributors}}]{astropy:2018}
{Astropy Collaboration}, {Price-Whelan}, A.~M., {Sip{\H o}cz}, B.~M., {et~al.}
  2018, \aj, 156, 123, \dodoi{10.3847/1538-3881/aabc4f}

\bibitem[{{Astropy Collaboration} {et~al.}(2022){Astropy Collaboration},
  {Price-Whelan}, {Lim}, {Earl}, {Starkman}, {Bradley}, {Shupe}, {Patil},
  {Corrales}, {Brasseur}, {N{\"o}the}, {Donath}, {Tollerud}, {Morris},
  {Ginsburg}, {Vaher}, {Weaver}, {Tocknell}, {Jamieson}, {van Kerkwijk},
  {Robitaille}, {Merry}, {Bachetti}, {G{\"u}nther}, {Aldcroft},
  {Alvarado-Montes}, {Archibald}, {B{\'o}di}, {Bapat}, {Barentsen},
  {Baz{\'a}n}, {Biswas}, {Boquien}, {Burke}, {Cara}, {Cara}, {Conroy},
  {Conseil}, {Craig}, {Cross}, {Cruz}, {D'Eugenio}, {Dencheva}, {Devillepoix},
  {Dietrich}, {Eigenbrot}, {Erben}, {Ferreira}, {Foreman-Mackey}, {Fox},
  {Freij}, {Garg}, {Geda}, {Glattly}, {Gondhalekar}, {Gordon}, {Grant},
  {Greenfield}, {Groener}, {Guest}, {Gurovich}, {Handberg}, {Hart},
  {Hatfield-Dodds}, {Homeier}, {Hosseinzadeh}, {Jenness}, {Jones}, {Joseph},
  {Kalmbach}, {Karamehmetoglu}, {Ka{\l}uszy{\'n}ski}, {Kelley}, {Kern},
  {Kerzendorf}, {Koch}, {Kulumani}, {Lee}, {Ly}, {Ma}, {MacBride}, {Maljaars},
  {Muna}, {Murphy}, {Norman}, {O'Steen}, {Oman}, {Pacifici}, {Pascual},
  {Pascual-Granado}, {Patil}, {Perren}, {Pickering}, {Rastogi}, {Roulston},
  {Ryan}, {Rykoff}, {Sabater}, {Sakurikar}, {Salgado}, {Sanghi}, {Saunders},
  {Savchenko}, {Schwardt}, {Seifert-Eckert}, {Shih}, {Jain}, {Shukla}, {Sick},
  {Simpson}, {Singanamalla}, {Singer}, {Singhal}, {Sinha}, {Sip{\H{o}}cz},
  {Spitler}, {Stansby}, {Streicher}, {{\v{S}}umak}, {Swinbank}, {Taranu},
  {Tewary}, {Tremblay}, {Val-Borro}, {Van Kooten}, {Vasovi{\'c}}, {Verma}, {de
  Miranda Cardoso}, {Williams}, {Wilson}, {Winkel}, {Wood-Vasey}, {Xue},
  {Yoachim}, {Zhang}, {Zonca}, \& {Astropy Project
  Contributors}}]{astropy:2022}
{Astropy Collaboration}, {Price-Whelan}, A.~M., {Lim}, P.~L., {et~al.} 2022,
  \apj, 935, 167, \dodoi{10.3847/1538-4357/ac7c74}

\bibitem[{{Bagley} {et~al.}(2023){Bagley}, {Finkelstein}, {Koekemoer},
  {Ferguson}, {Arrabal Haro}, {Dickinson}, {Kartaltepe}, {Papovich},
  {P{\'e}rez-Gonz{\'a}lez}, {Pirzkal}, {Somerville}, {Willmer}, {Yang}, {Yung},
  {Fontana}, {Grazian}, {Grogin}, {Hirschmann}, {Kewley}, {Kirkpatrick},
  {Kocevski}, {Lotz}, {Medrano}, {Morales}, {Pentericci}, {Ravindranath},
  {Trump}, {Wilkins}, {Calabr{\`o}}, {Cooper}, {Costantin}, {de la Vega},
  {Hilbert}, {Hutchison}, {Larson}, {Lucas}, {McGrath}, {Ryan}, {Wang}, \&
  {Wuyts}}]{Bagley2023}
{Bagley}, M.~B., {Finkelstein}, S.~L., {Koekemoer}, A.~M., {et~al.} 2023,
  \apjl, 946, L12, \dodoi{10.3847/2041-8213/acbb08}

\bibitem[{{Bertin} \& {Arnouts}(1996)}]{sourceextractor}
{Bertin}, E., \& {Arnouts}, S. 1996, \aaps, 117, 393,
  \dodoi{10.1051/aas:1996164}

\bibitem[{{Boquien} {et~al.}(2019){Boquien}, {Burgarella}, {Roehlly}, {Buat},
  {Ciesla}, {Corre}, {Inoue}, \& {Salas}}]{cigale2}
{Boquien}, M., {Burgarella}, D., {Roehlly}, Y., {et~al.} 2019, \aap, 622, A103,
  \dodoi{10.1051/0004-6361/201834156}

\bibitem[{{Boylan-Kolchin}(2022)}]{Boylan-Kolchin2022}
{Boylan-Kolchin}, M. 2022, arXiv e-prints, arXiv:2208.01611.
\newblock \doarXiv{2208.01611}

\bibitem[{{Brammer} {et~al.}(2008){Brammer}, {van Dokkum}, \&
  {Coppi}}]{Brammer2008}
{Brammer}, G.~B., {van Dokkum}, P.~G., \& {Coppi}, P. 2008, \apj, 686, 1503,
  \dodoi{10.1086/591786}

\bibitem[{{Brammer} {et~al.}(2011){Brammer}, {Whitaker}, {van Dokkum},
  {Marchesini}, {Franx}, {Kriek}, {Labb{\'e}}, {Lee}, {Muzzin}, {Quadri},
  {Rudnick}, \& {Williams}}]{Brammer2011}
{Brammer}, G.~B., {Whitaker}, K.~E., {van Dokkum}, P.~G., {et~al.} 2011, \apj,
  739, 24, \dodoi{10.1088/0004-637X/739/1/24}

\bibitem[{{Bruzual} \& {Charlot}(2003)}]{bruz03}
{Bruzual}, G., \& {Charlot}, S. 2003, \mnras, 344, 1000,
  \dodoi{10.1046/j.1365-8711.2003.06897.x}

\bibitem[{{Bunker} {et~al.}(2020){Bunker}, {NIRSPEC Instrument Science Team},
  \& {JAESs Collaboration}}]{Bunker2020}
{Bunker}, A.~J., {NIRSPEC Instrument Science Team}, \& {JAESs Collaboration}.
  2020, in Uncovering Early Galaxy Evolution in the ALMA and JWST Era, ed.
  E.~{da Cunha}, J.~{Hodge}, J.~{Afonso}, L.~{Pentericci}, \& D.~{Sobral}, Vol.
  352, 342--346, \dodoi{10.1017/S1743921319009463}

\bibitem[{{Burgarella} {et~al.}(2005){Burgarella}, {Buat}, \&
  {Iglesias-P{\'a}ramo}}]{cigale}
{Burgarella}, D., {Buat}, V., \& {Iglesias-P{\'a}ramo}, J. 2005, \mnras, 360,
  1413, \dodoi{10.1111/j.1365-2966.2005.09131.x}

\bibitem[{{Calzetti} {et~al.}(2000){Calzetti}, {Armus}, {Bohlin}, {Kinney},
  {Koornneef}, \& {Storchi-Bergmann}}]{calzetti00}
{Calzetti}, D., {Armus}, L., {Bohlin}, R.~C., {et~al.} 2000, \apj, 533, 682,
  \dodoi{10.1086/308692}

\bibitem[{{Carnall} {et~al.}(2020){Carnall}, {Walker}, {McLure}, {Dunlop},
  {McLeod}, {Cullen}, {Wild}, {Amorin}, {Bolzonella}, {Castellano}, {Cimatti},
  {Cucciati}, {Fontana}, {Gargiulo}, {Garilli}, {Jarvis}, {Pentericci},
  {Pozzetti}, {Zamorani}, {Calabro}, {Hathi}, \& {Koekemoer}}]{Carnall2020}
{Carnall}, A.~C., {Walker}, S., {McLure}, R.~J., {et~al.} 2020, \mnras, 496,
  695, \dodoi{10.1093/mnras/staa1535}

\bibitem[{{Carnall} {et~al.}(2023){Carnall}, {McLeod}, {McLure}, {Dunlop},
  {Begley}, {Cullen}, {Donnan}, {Hamadouche}, {Jewell}, {Jones}, {Pollock}, \&
  {Wild}}]{Carnall2023}
{Carnall}, A.~C., {McLeod}, D.~J., {McLure}, R.~J., {et~al.} 2023, \mnras,
  \dodoi{10.1093/mnras/stad369}

\bibitem[{Casey {et~al.}(2014)Casey, Narayanan, \& Cooray}]{casey2014dusty}
Casey, C.~M., Narayanan, D., \& Cooray, A. 2014, Physics Reports, 541, 45

\bibitem[{{Casey} {et~al.}(2022){Casey}, {Kartaltepe}, {Drakos}, {Franco},
  {Ilbert}, {Rose}, {Cox}, {Nightingale}, {Robertson}, {Silverman},
  {Koekemoer}, {Massey}, {McCracken}, {Rhodes}, {Akins}, {Amvrosiadis},
  {Arango-Toro}, {Bagley}, {Capak}, {Champagne}, {Chartab}, {Chavez Ortiz},
  {Cooke}, {Cooper}, {Darvish}, {Ding}, {Faisst}, {Finkelstein}, {Fujimoto},
  {Gentile}, {Gillman}, {Gould}, {Gozaliasl}, {Harish}, {Hayward}, {He},
  {Hemmati}, {Hirschmann}, {Jin}, {Khostovan}, {Kokorev}, {Lambrides},
  {Laigle}, {Leung}, {Liu}, {Liaudat}, {Long}, {Magdis}, {Mahler}, {Mainieri},
  {Manning}, {Maraston}, {Martin}, {McCleary}, {McKinney}, {McPartland},
  {Mobasher}, {Pattnaik}, {Renzini}, {Rich}, {Sanders}, {Sattari},
  {Scognamiglio}, {Scoville}, {Sheth}, {Shuntov}, {Sparre}, {Suzuki}, {Talia},
  {Toft}, {Trakhtenbrot}, {Urry}, {Valentino}, {Vanderhoof}, {Vardoulaki},
  {Weaver}, {Whitaker}, {Wilkins}, {Yang}, \& {Zavala}}]{Casey2022}
{Casey}, C.~M., {Kartaltepe}, J.~S., {Drakos}, N.~E., {et~al.} 2022, arXiv
  e-prints, arXiv:2211.07865.
\newblock \doarXiv{2211.07865}

\bibitem[{{Cecchi} {et~al.}(2019){Cecchi}, {Bolzonella}, {Cimatti}, \&
  {Girelli}}]{Cecchi2019}
{Cecchi}, R., {Bolzonella}, M., {Cimatti}, A., \& {Girelli}, G. 2019, \apjl,
  880, L14, \dodoi{10.3847/2041-8213/ab2c80}

\bibitem[{{Chabrier}(2003)}]{chabrier}
{Chabrier}, G. 2003, \pasp, 115, 763, \dodoi{10.1086/376392}

\bibitem[{{Chevallard} \& {Charlot}(2016)}]{Chevallard2016}
{Chevallard}, J., \& {Charlot}, S. 2016, \mnras, 462, 1415,
  \dodoi{10.1093/mnras/stw1756}

\bibitem[{{Conroy} \& {Gunn}(2010)}]{Conroy2010}
{Conroy}, C., \& {Gunn}, J.~E. 2010, \apj, 712, 833,
  \dodoi{10.1088/0004-637X/712/2/833}

\bibitem[{{da Cunha} {et~al.}(2015){da Cunha}, {Walter}, {Smail}, {Swinbank},
  {Simpson}, {Decarli}, {Hodge}, {Weiss}, {van der Werf}, {Bertoldi},
  {Chapman}, {Cox}, {Danielson}, {Dannerbauer}, {Greve}, {Ivison}, {Karim}, \&
  {Thomson}}]{dacunha15}
{da Cunha}, E., {Walter}, F., {Smail}, I.~R., {et~al.} 2015, \apj, 806, 110,
  \dodoi{10.1088/0004-637X/806/1/110}

\bibitem[{{Deshmukh} {et~al.}(2018){Deshmukh}, {Caputi}, {Ashby}, {Cowley},
  {McCracken}, {Fynbo}, {Le F{\`e}vre}, {Milvang-Jensen}, \&
  {Ilbert}}]{deshmukh18}
{Deshmukh}, S., {Caputi}, K.~I., {Ashby}, M.~L.~N., {et~al.} 2018, \apj, 864,
  166, \dodoi{10.3847/1538-4357/aad9f5}

\bibitem[{{D'Eugenio} {et~al.}(2020){D'Eugenio}, {Daddi}, {Gobat},
  {Strazzullo}, {Lustig}, {Delvecchio}, {Jin}, {Puglisi}, {Calabr{\'o}},
  {Mancini}, {Dickinson}, {Cimatti}, \& {Onodera}}]{D'Eugenio2020}
{D'Eugenio}, C., {Daddi}, E., {Gobat}, R., {et~al.} 2020, \apjl, 892, L2,
  \dodoi{10.3847/2041-8213/ab7a96}

\bibitem[{{Drakos} {et~al.}(2022){Drakos}, {Villasenor}, {Robertson}, {Hausen},
  {Dickinson}, {Ferguson}, {Furlanetto}, {Greene}, {Madau}, {Shapley}, {Stark},
  \& {Wechsler}}]{Drakos2022}
{Drakos}, N.~E., {Villasenor}, B., {Robertson}, B.~E., {et~al.} 2022, \apj,
  926, 194, \dodoi{10.3847/1538-4357/ac46fb}

\bibitem[{{Dunlop} {et~al.}(2021){Dunlop}, {Abraham}, {Ashby}, {Bagley},
  {Best}, {Bongiorno}, {Bouwens}, {Bowler}, {Brammer}, {Bremer}, {Calabro'},
  {Carnall}, {Castellano}, {Cirasuolo}, {Conselice}, {Cullen}, {Dave}, {Dayal},
  {Dekel}, {Dickinson}, {Duncan}, {Elbaz}, {Ellis}, {Ferguson}, {Ferrara},
  {Finkelstein}, {Fontana}, {Furlanetto}, {Fynbo}, {Gallerani}, {Gardner},
  {Giavalisco}, {Grazian}, {Grogin}, {Harikane}, {Hopkins}, {Ilbert},
  {Illingworth}, {Juneau}, {Jung}, {Kartaltepe}, {Kassin}, {Kauffmann},
  {Khochfar}, {Kirkpatrick}, {Kocevski}, {Koekemoer}, {Labbe}, {Laporte},
  {Larson}, {Lucas}, {Magee}, {Mason}, {McCracken}, {McLeod}, {McLure},
  {Merlin}, {Mesinger}, {Milvang-Jensen}, {Newman}, {Oesch}, {Ouchi},
  {Pacifici}, {Papovich}, {Peacock}, {Peeples}, {Pentericci}, {Perez-Gonzalez},
  {Pirzkal}, {Pope}, {Pye}, {Reddy}, {Robertson}, {Salvato}, {Santini},
  {Schaerer}, {Shapley}, {Simons}, {Smit}, {Smith}, {Snyder}, {Somerville},
  {Stanway}, {Stefanon}, {Tasca}, {Tikkanen}, {Tresse}, {Trump}, {Whitaker},
  {Wilkins}, {Wright}, {Wyithe}, {van Dokkum}, \& {van der Werf}}]{Dunlop2021}
{Dunlop}, J.~S., {Abraham}, R.~G., {Ashby}, M. L.~N., {et~al.} 2021, {PRIMER:
  Public Release IMaging for Extragalactic Research}, JWST Proposal. Cycle 1,
  ID. \#1837

\bibitem[{{Finkelstein} {et~al.}(2017){Finkelstein}, {Dickinson}, {Ferguson},
  {Grazian}, {Grogin}, {Kartaltepe}, {Kewley}, {Kocevski}, {Koekemoer}, {Lotz},
  {Papovich}, {Pentericci}, {Perez-Gonzalez}, {Pirzkal}, {Ravindranath},
  {Somerville}, {Trump}, \& {Wilkins}}]{Finkelstein2017}
{Finkelstein}, S.~L., {Dickinson}, M., {Ferguson}, H.~C., {et~al.} 2017, {The
  Cosmic Evolution Early Release Science (CEERS) Survey}, JWST Proposal ID
  1345. Cycle 0 Early Release Science

\bibitem[{{Finkelstein} {et~al.}(2023){Finkelstein}, {Bagley}, {Ferguson},
  {Wilkins}, {Kartaltepe}, {Papovich}, {Yung}, {Arrabal Haro}, {Behroozi},
  {Dickinson}, {Kocevski}, {Koekemoer}, {Larson}, {Le Bail}, {Morales},
  {P{\'e}rez-Gonz{\'a}lez}, {Burgarella}, {Dav{\'e}}, {Hirschmann},
  {Somerville}, {Wuyts}, {Bromm}, {Casey}, {Fontana}, {Fujimoto}, {Gardner},
  {Giavalisco}, {Grazian}, {Grogin}, {Hathi}, {Hutchison}, {Jha}, {Jogee},
  {Kewley}, {Kirkpatrick}, {Long}, {Lotz}, {Pentericci}, {Pierel}, {Pirzkal},
  {Ravindranath}, {Ryan}, {Trump}, {Yang}, {Bhatawdekar}, {Bisigello}, {Buat},
  {Calabr{\`o}}, {Castellano}, {Cleri}, {Cooper}, {Croton}, {Daddi}, {Dekel},
  {Elbaz}, {Franco}, {Gawiser}, {Holwerda}, {Huertas-Company}, {Jaskot},
  {Leung}, {Lucas}, {Mobasher}, {Pandya}, {Tacchella}, {Weiner}, \&
  {Zavala}}]{Finkelstein2023}
{Finkelstein}, S.~L., {Bagley}, M.~B., {Ferguson}, H.~C., {et~al.} 2023, \apjl,
  946, L13, \dodoi{10.3847/2041-8213/acade4}

\bibitem[{{Forrest} {et~al.}(2020){Forrest}, {Marsan}, {Annunziatella},
  {Wilson}, {Muzzin}, {Marchesini}, {Cooper}, {Chan}, {McConachie}, {Gomez},
  {Kado-Fong}, {Barbera}, {Lange-Vagle}, {Nantais}, {Nonino}, {Saracco},
  {Stefanon}, \& {van der Burg}}]{Forrest2020}
{Forrest}, B., {Marsan}, Z.~C., {Annunziatella}, M., {et~al.} 2020, \apj, 903,
  47, \dodoi{10.3847/1538-4357/abb819}

\bibitem[{{Girelli} {et~al.}(2019){Girelli}, {Bolzonella}, \&
  {Cimatti}}]{Girelli2019}
{Girelli}, G., {Bolzonella}, M., \& {Cimatti}, A. 2019, \aap, 632, A80,
  \dodoi{10.1051/0004-6361/201834547}

\bibitem[{{Glazebrook} {et~al.}(2017){Glazebrook}, {Schreiber}, {Labb{\'e}},
  {Nanayakkara}, {Kacprzak}, {Oesch}, {Papovich}, {Spitler}, {Straatman},
  {Tran}, \& {Yuan}}]{Glazebrook2017}
{Glazebrook}, K., {Schreiber}, C., {Labb{\'e}}, I., {et~al.} 2017, \nat, 544,
  71, \dodoi{10.1038/nature21680}

\bibitem[{{Gould} {et~al.}(2023){Gould}, {Brammer}, {Valentino}, {Whitaker},
  {Weaver}, {Lagos}, {Rizzo}, {Franco}, {Hseih}, {Ilbert}, {Jin}, {Magdis},
  {McCracken}, {Mobasher}, {Shuntov}, {Steinhardt}, {Strait}, \&
  {Toft}}]{Gould2023}
{Gould}, K. M.~L., {Brammer}, G., {Valentino}, F., {et~al.} 2023, arXiv
  e-prints, arXiv:2302.10934, \dodoi{10.48550/arXiv.2302.10934}

\bibitem[{{Grogin} {et~al.}(2011){Grogin}, {Kocevski}, {Faber}, {Ferguson},
  {Koekemoer}, {Riess}, {Acquaviva}, {Alexander}, {Almaini}, {Ashby}, {Barden},
  {Bell}, {Bournaud}, {Brown}, {Caputi}, {Casertano}, {Cassata}, {Castellano},
  {Challis}, {Chary}, {Cheung}, {Cirasuolo}, {Conselice}, {Roshan Cooray},
  {Croton}, {Daddi}, {Dahlen}, {Dav{\'e}}, {de Mello}, {Dekel}, {Dickinson},
  {Dolch}, {Donley}, {Dunlop}, {Dutton}, {Elbaz}, {Fazio}, {Filippenko},
  {Finkelstein}, {Fontana}, {Gardner}, {Garnavich}, {Gawiser}, {Giavalisco},
  {Grazian}, {Guo}, {Hathi}, {H{\"a}ussler}, {Hopkins}, {Huang}, {Huang},
  {Jha}, {Kartaltepe}, {Kirshner}, {Koo}, {Lai}, {Lee}, {Li}, {Lotz}, {Lucas},
  {Madau}, {McCarthy}, {McGrath}, {McIntosh}, {McLure}, {Mobasher},
  {Moustakas}, {Mozena}, {Nandra}, {Newman}, {Niemi}, {Noeske}, {Papovich},
  {Pentericci}, {Pope}, {Primack}, {Rajan}, {Ravindranath}, {Reddy}, {Renzini},
  {Rix}, {Robaina}, {Rodney}, {Rosario}, {Rosati}, {Salimbeni}, {Scarlata},
  {Siana}, {Simard}, {Smidt}, {Somerville}, {Spinrad}, {Straughn}, {Strolger},
  {Telford}, {Teplitz}, {Trump}, {van der Wel}, {Villforth}, {Wechsler},
  {Weiner}, {Wiklind}, {Wild}, {Wilson}, {Wuyts}, {Yan}, \& {Yun}}]{Grogin2011}
{Grogin}, N.~A., {Kocevski}, D.~D., {Faber}, S.~M., {et~al.} 2011, \apjs, 197,
  35, \dodoi{10.1088/0067-0049/197/2/35}

\bibitem[{{Hatziminaoglou} {et~al.}(2005){Hatziminaoglou}, {P{\'e}rez-Fournon},
  {Polletta}, {Afonso-Luis}, {Hern{\'a}n-Caballero}, {Montenegro-Montes},
  {Lonsdale}, {Xu}, {Franceschini}, {Rowan-Robinson}, {Babbedge}, {Smith},
  {Surace}, {Shupe}, {Fang}, {Farrah}, {Oliver}, {Gonz{\'a}lez-Solares}, \&
  {Serjeant}}]{Hatziminaoglou2005}
{Hatziminaoglou}, E., {P{\'e}rez-Fournon}, I., {Polletta}, M., {et~al.} 2005,
  \aj, 129, 1198, \dodoi{10.1086/428003}

\bibitem[{{Hwang} {et~al.}(2021){Hwang}, {Wang}, {Chang}, {Lim}, {Chen}, {Gao},
  {Dunlop}, {Gao}, {Ho}, {Hwang}, {Koprowski}, {Micha{\l}owski}, {Peng},
  {Shim}, {Simpson}, \& {Toba}}]{Hwang2021}
{Hwang}, Y.-H., {Wang}, W.-H., {Chang}, Y.-Y., {et~al.} 2021, \apj, 913, 6,
  \dodoi{10.3847/1538-4357/abf11a}

\bibitem[{{Koekemoer} {et~al.}(2011){Koekemoer}, {Faber}, {Ferguson}, {Grogin},
  {Kocevski}, {Koo}, {Lai}, {Lotz}, {Lucas}, {McGrath}, {Ogaz}, {Rajan},
  {Riess}, {Rodney}, {Strolger}, {Casertano}, {Castellano}, {Dahlen},
  {Dickinson}, {Dolch}, {Fontana}, {Giavalisco}, {Grazian}, {Guo}, {Hathi},
  {Huang}, {van der Wel}, {Yan}, {Acquaviva}, {Alexander}, {Almaini}, {Ashby},
  {Barden}, {Bell}, {Bournaud}, {Brown}, {Caputi}, {Cassata}, {Challis},
  {Chary}, {Cheung}, {Cirasuolo}, {Conselice}, {Roshan Cooray}, {Croton},
  {Daddi}, {Dav{\'e}}, {de Mello}, {de Ravel}, {Dekel}, {Donley}, {Dunlop},
  {Dutton}, {Elbaz}, {Fazio}, {Filippenko}, {Finkelstein}, {Frazer}, {Gardner},
  {Garnavich}, {Gawiser}, {Gruetzbauch}, {Hartley}, {H{\"a}ussler},
  {Herrington}, {Hopkins}, {Huang}, {Jha}, {Johnson}, {Kartaltepe},
  {Khostovan}, {Kirshner}, {Lani}, {Lee}, {Li}, {Madau}, {McCarthy},
  {McIntosh}, {McLure}, {McPartland}, {Mobasher}, {Moreira}, {Mortlock},
  {Moustakas}, {Mozena}, {Nandra}, {Newman}, {Nielsen}, {Niemi}, {Noeske},
  {Papovich}, {Pentericci}, {Pope}, {Primack}, {Ravindranath}, {Reddy},
  {Renzini}, {Rix}, {Robaina}, {Rosario}, {Rosati}, {Salimbeni}, {Scarlata},
  {Siana}, {Simard}, {Smidt}, {Snyder}, {Somerville}, {Spinrad}, {Straughn},
  {Telford}, {Teplitz}, {Trump}, {Vargas}, {Villforth}, {Wagner}, {Wandro},
  {Wechsler}, {Weiner}, {Wiklind}, {Wild}, {Wilson}, {Wuyts}, \&
  {Yun}}]{Koekemoer2011}
{Koekemoer}, A.~M., {Faber}, S.~M., {Ferguson}, H.~C., {et~al.} 2011, \apjs,
  197, 36, \dodoi{10.1088/0067-0049/197/2/36}

\bibitem[{{Labb{\'e}} {et~al.}(2005){Labb{\'e}}, {Huang}, {Franx}, {Rudnick},
  {Barmby}, {Daddi}, {van Dokkum}, {Fazio}, {F{\"o}rster Schreiber},
  {Moorwood}, {Rix}, {R{\"o}ttgering}, {Trujillo}, \& {van der
  Werf}}]{Labbe2005}
{Labb{\'e}}, I., {Huang}, J., {Franx}, M., {et~al.} 2005, \apjl, 624, L81,
  \dodoi{10.1086/430700}

\bibitem[{{Labb{\'e}} {et~al.}(2013){Labb{\'e}}, {Oesch}, {Bouwens},
  {Illingworth}, {Magee}, {Gonz{\'a}lez}, {Carollo}, {Franx}, {Trenti}, {van
  Dokkum}, \& {Stiavelli}}]{Labbe2013}
{Labb{\'e}}, I., {Oesch}, P.~A., {Bouwens}, R.~J., {et~al.} 2013, \apjl, 777,
  L19, \dodoi{10.1088/2041-8205/777/2/L19}

\bibitem[{{Labbe} {et~al.}(2022){Labbe}, {van Dokkum}, {Nelson}, {Bezanson},
  {Suess}, {Leja}, {Brammer}, {Whitaker}, {Mathews}, \& {Stefanon}}]{Labbe2022}
{Labbe}, I., {van Dokkum}, P., {Nelson}, E., {et~al.} 2022, arXiv e-prints,
  arXiv:2207.12446.
\newblock \doarXiv{2207.12446}

\bibitem[{{Larson} {et~al.}(2022){Larson}, {Hutchison}, {Bagley},
  {Finkelstein}, {Yung}, {Somerville}, {Hirschmann}, {Brammer}, {Holwerda},
  {Papovich}, {Morales}, \& {Wilkins}}]{Larson2022}
{Larson}, R.~L., {Hutchison}, T.~A., {Bagley}, M., {et~al.} 2022, arXiv
  e-prints, arXiv:2211.10035.
\newblock \doarXiv{2211.10035}

\bibitem[{{Liu} {et~al.}(2018){Liu}, {Jia}, {Yesuf}, {Faber}, {Koo}, {Guo},
  {Bell}, {Jiang}, {Wang}, {Koekemoer}, {Zheng}, {Fang}, {Barro},
  {P{\'e}rez-Gonz{\'a}lez}, {Dekel}, {Kocevski}, {Hathi}, {Croton},
  {Huertas-Company}, {Meng}, {Tong}, \& {Liu}}]{Liu2018}
{Liu}, F.~S., {Jia}, M., {Yesuf}, H.~M., {et~al.} 2018, \apj, 860, 60,
  \dodoi{10.3847/1538-4357/aac20d}

\bibitem[{{Long} {et~al.}(2022){Long}, {Casey}, {Lagos}, {Lambrides}, {Zavala},
  {Champagne}, {Cooper}, \& {Cooray}}]{Long2022}
{Long}, A.~S., {Casey}, C.~M., {Lagos}, C. d.~P., {et~al.} 2022, arXiv
  e-prints, arXiv:2211.02072.
\newblock \doarXiv{2211.02072}

\bibitem[{{Lovell} {et~al.}(2023){Lovell}, {Harrison}, {Harikane}, {Tacchella},
  \& {Wilkins}}]{Lovell2023}
{Lovell}, C.~C., {Harrison}, I., {Harikane}, Y., {Tacchella}, S., \& {Wilkins},
  S.~M. 2023, \mnras, 518, 2511, \dodoi{10.1093/mnras/stac3224}

\bibitem[{{Lovell} {et~al.}(2021){Lovell}, {Vijayan}, {Thomas}, {Wilkins},
  {Barnes}, {Irodotou}, \& {Roper}}]{Lovell2021a}
{Lovell}, C.~C., {Vijayan}, A.~P., {Thomas}, P.~A., {et~al.} 2021, \mnras, 500,
  2127, \dodoi{10.1093/mnras/staa3360}

\bibitem[{{Lovell} {et~al.}(2022){Lovell}, {Roper}, {Vijayan}, {Seeyave},
  {Irodotou}, {Wilkins}, {Conselice}, {Fortuni}, {Kuusisto}, {Merlin},
  {Santini}, \& {Thomas}}]{Lovell2022}
{Lovell}, C.~C., {Roper}, W., {Vijayan}, A.~P., {et~al.} 2022, arXiv e-prints,
  arXiv:2211.07540.
\newblock \doarXiv{2211.07540}

\bibitem[{{Marsan} {et~al.}(2022){Marsan}, {Muzzin}, {Marchesini}, {Stefanon},
  {Martis}, {Annunziatella}, {Chan}, {Cooper}, {Forrest}, {Gomez},
  {McConachie}, \& {Wilson}}]{Marsan2022}
{Marsan}, Z.~C., {Muzzin}, A., {Marchesini}, D., {et~al.} 2022, \apj, 924, 25,
  \dodoi{10.3847/1538-4357/ac312a}

\bibitem[{{Martis} {et~al.}(2019){Martis}, {Marchesini}, {Muzzin}, {Stefanon},
  {Brammer}, {da Cunha}, {Sajina}, \& {Labbe}}]{Martis2019}
{Martis}, N.~S., {Marchesini}, D.~M., {Muzzin}, A., {et~al.} 2019, \apj, 882,
  65, \dodoi{10.3847/1538-4357/ab32f1}

\bibitem[{{McKinney} {et~al.}(2023){McKinney}, {Finnerty}, {Casey}, {Franco},
  {Long}, {Fujimoto}, {Zavala}, {Cooper}, {Akins}, {Pope}, {Armus}, {Soifer},
  {Larson}, {Matthews}, {Melbourne}, \& {Cushing}}]{McKinney2023}
{McKinney}, J., {Finnerty}, L., {Casey}, C.~M., {et~al.} 2023, \apjl, 946, L39,
  \dodoi{10.3847/2041-8213/acc322}

\bibitem[{{Merlin} {et~al.}(2018){Merlin}, {Fontana}, {Castellano}, {Santini},
  {Torelli}, {Boutsia}, {Wang}, {Grazian}, {Pentericci}, {Schreiber}, {Ciesla},
  {McLure}, {Derriere}, {Dunlop}, \& {Elbaz}}]{Merlin2018}
{Merlin}, E., {Fontana}, A., {Castellano}, M., {et~al.} 2018, \mnras, 473,
  2098, \dodoi{10.1093/mnras/stx2385}

\bibitem[{{Merlin} {et~al.}(2019){Merlin}, {Fortuni}, {Torelli}, {Santini},
  {Castellano}, {Fontana}, {Grazian}, {Pentericci}, {Pilo}, \&
  {Schmidt}}]{Merlin2019}
{Merlin}, E., {Fortuni}, F., {Torelli}, M., {et~al.} 2019, \mnras, 490, 3309,
  \dodoi{10.1093/mnras/stz2615}

\bibitem[{{Muzzin} {et~al.}(2013){Muzzin}, {Marchesini}, {Stefanon}, {Franx},
  {McCracken}, {Milvang-Jensen}, {Dunlop}, {Fynbo}, {Brammer}, {Labb{\'e}}, \&
  {van Dokkum}}]{muzzin13}
{Muzzin}, A., {Marchesini}, D., {Stefanon}, M., {et~al.} 2013, \apj, 777, 18,
  \dodoi{10.1088/0004-637X/777/1/18}

\bibitem[{{Nanayakkara} {et~al.}(2022){Nanayakkara}, {Glazebrook}, {Jacobs},
  {Schreiber}, {Brammer}, {Esdaile}, {Kacprzak}, {Labbe}, {Lagos},
  {Marchesini}, {Marsan}, {Nateghi}, {Oesch}, {Papovich}, {Remus}, \&
  {Tran}}]{Nanayakkara2022}
{Nanayakkara}, T., {Glazebrook}, K., {Jacobs}, C., {et~al.} 2022, arXiv
  e-prints, arXiv:2212.11638, \dodoi{10.48550/arXiv.2212.11638}

\bibitem[{{Noeske} {et~al.}(2007){Noeske}, {Weiner}, {Faber}, {Papovich},
  {Koo}, {Somerville}, {Bundy}, {Conselice}, {Newman}, {Schiminovich}, {Le
  Floc'h}, {Coil}, {Rieke}, {Lotz}, {Primack}, {Barmby}, {Cooper}, {Davis},
  {Ellis}, {Fazio}, {Guhathakurta}, {Huang}, {Kassin}, {Martin}, {Phillips},
  {Rich}, {Small}, {Willmer}, \& {Wilson}}]{Noeske2007a}
{Noeske}, K.~G., {Weiner}, B.~J., {Faber}, S.~M., {et~al.} 2007, \apjl, 660,
  L43, \dodoi{10.1086/517926}

\bibitem[{{Noll} {et~al.}(2009){Noll}, {Burgarella}, {Giovannoli}, {Buat},
  {Marcillac}, \& {Mu{\~n}oz-Mateos}}]{cigale3}
{Noll}, S., {Burgarella}, D., {Giovannoli}, E., {et~al.} 2009, \aap, 507, 1793,
  \dodoi{10.1051/0004-6361/200912497}

\bibitem[{{Pacifici} {et~al.}(2016){Pacifici}, {Kassin}, {Weiner}, {Holden},
  {Gardner}, {Faber}, {Ferguson}, {Koo}, {Primack}, {Bell}, {Dekel}, {Gawiser},
  {Giavalisco}, {Rafelski}, {Simons}, {Barro}, {Croton}, {Dav{\'e}}, {Fontana},
  {Grogin}, {Koekemoer}, {Lee}, {Salmon}, {Somerville}, \&
  {Behroozi}}]{Pacifici2016}
{Pacifici}, C., {Kassin}, S.~A., {Weiner}, B.~J., {et~al.} 2016, \apj, 832, 79,
  \dodoi{10.3847/0004-637X/832/1/79}

\bibitem[{{P{\'e}rez-Gonz{\'a}lez} {et~al.}(2022){P{\'e}rez-Gonz{\'a}lez},
  {Barro}, {Annunziatella}, {Costantin}, {Garc{\'\i}a-Argum{\'a}nez},
  {McGrath}, {M{\'e}rida}, {Zavala}, {Arrabal Haro}, {Bagley}, {Backhaus},
  {Behroozi}, {Bell}, {Buat}, {Calabr{\`o}}, {Casey}, {Cleri}, {Coogan},
  {Cooper}, {Cooray}, {Dekel}, {Dickinson}, {Elbaz}, {Ferguson}, {Finkelstein},
  {Fontana}, {Franco}, {Gardner}, {Giavalisco}, {G{\'o}mez-Guijarro},
  {Grazian}, {Grogin}, {Guo}, {Jogee}, {Kartaltepe}, {Kewley}, {Kirkpatrick},
  {Kocevski}, {Koekemoer}, {Long}, {Lotz}, {Lucas}, {Papovich}, {Pirzkal},
  {Ravindranath}, {Somerville}, {Tacchella}, {Trump}, {Wang}, {Wilkins},
  {Wuyts}, {Yang}, \& {Yung}}]{Perez-Gonzalez2022}
{P{\'e}rez-Gonz{\'a}lez}, P.~G., {Barro}, G., {Annunziatella}, M., {et~al.}
  2022, arXiv e-prints, arXiv:2211.00045.
\newblock \doarXiv{2211.00045}

\bibitem[{{Planck Collaboration} {et~al.}(2016){Planck Collaboration}, {Ade},
  {Aghanim}, {Arnaud}, {Ashdown}, {Aumont}, {Baccigalupi}, {Banday},
  {Barreiro}, {Bartlett}, \& et~al.}]{PlanckCollaboration2016}
{Planck Collaboration}, {Ade}, P.~A.~R., {Aghanim}, N., {et~al.} 2016, \aap,
  594, A13, \dodoi{10.1051/0004-6361/201525830}

\bibitem[{{Robertson} {et~al.}(2022){Robertson}, {Tacchella}, {Johnson},
  {Hainline}, {Whitler}, {Eisenstein}, {Endsley}, {Rieke}, {Stark}, {Alberts},
  {Dressler}, {Egami}, {Hausen}, {Rieke}, {Shivaei}, {Williams}, {Willmer},
  {Arribas}, {Bonaventura}, {Bunker}, {Cameron}, {Carniani}, {Charlot},
  {Chevallard}, {Curti}, {Curtis-Lake}, {D'Eugenio}, {Jakobsen}, {Looser},
  {L{\"u}tzgendorf}, {Maiolino}, {Maseda}, {Rawle}, {Rix}, {Smit}, {{\"U}bler},
  {Willott}, {Witstok}, {Baum}, {Bhatawdekar}, {Boyett}, {Chen}, {de Graaff},
  {Florian}, {Helton}, {Hviding}, {Ji}, {Kumari}, {Lyu}, {Nelson}, {Sandles},
  {Saxena}, {Suess}, {Sun}, {Topping}, \& {Wallace}}]{Robertson2022}
{Robertson}, B.~E., {Tacchella}, S., {Johnson}, B.~D., {et~al.} 2022, arXiv
  e-prints, arXiv:2212.04480.
\newblock \doarXiv{2212.04480}

\bibitem[{{Santini} {et~al.}(2019){Santini}, {Merlin}, {Fontana}, {Magnelli},
  {Paris}, {Castellano}, {Grazian}, {Pentericci}, {Pilo}, \&
  {Torelli}}]{Santini2019}
{Santini}, P., {Merlin}, E., {Fontana}, A., {et~al.} 2019, \mnras, 486, 560,
  \dodoi{10.1093/mnras/stz801}

\bibitem[{{Schreiber} {et~al.}(2018){Schreiber}, {Glazebrook}, {Nanayakkara},
  {Kacprzak}, {Labb{\'e}}, {Oesch}, {Yuan}, {Tran}, {Papovich}, {Spitler}, \&
  {Straatman}}]{Schreiber2018}
{Schreiber}, C., {Glazebrook}, K., {Nanayakkara}, T., {et~al.} 2018, \aap, 618,
  A85, \dodoi{10.1051/0004-6361/201833070}

\bibitem[{{Shahidi} {et~al.}(2020){Shahidi}, {Mobasher}, {Nayyeri}, {Hemmati},
  {Wiklind}, {Chartab}, {Dickinson}, {Finkelstein}, {Pacifici}, {Papovich},
  {Ferguson}, {Fontana}, {Giavalisco}, {Koekemoer}, {Newman}, {Sattari}, \&
  {Somerville}}]{Shahidi2020}
{Shahidi}, A., {Mobasher}, B., {Nayyeri}, H., {et~al.} 2020, \apj, 897, 44,
  \dodoi{10.3847/1538-4357/ab96c5}

\bibitem[{{Simpson} {et~al.}(2014){Simpson}, {Swinbank}, {Smail}, {Alexand er},
  {Brandt}, {Bertoldi}, {de Breuck}, {Chapman}, {Coppin}, {da Cunha},
  {Danielson}, {Dannerbauer}, {Greve}, {Hodge}, {Ivison}, {Karim}, {Knudsen},
  {Poggianti}, {Schinnerer}, {Thomson}, {Walter}, {Wardlow}, {Wei{\ss}}, \&
  {van der Werf}}]{simpson14}
{Simpson}, J.~M., {Swinbank}, A.~M., {Smail}, I., {et~al.} 2014, \apj, 788,
  125, \dodoi{10.1088/0004-637X/788/2/125}

\bibitem[{{Smit} {et~al.}(2016){Smit}, {Bouwens}, {Labb{\'e}}, {Franx},
  {Wilkins}, \& {Oesch}}]{smit16}
{Smit}, R., {Bouwens}, R.~J., {Labb{\'e}}, I., {et~al.} 2016, \apj, 833, 254,
  \dodoi{10.3847/1538-4357/833/2/254}

\bibitem[{{Somerville} {et~al.}(2015){Somerville}, {Popping}, \&
  {Trager}}]{Somerville2015b}
{Somerville}, R.~S., {Popping}, G., \& {Trager}, S.~C. 2015, \mnras, 453, 4337,
  \dodoi{10.1093/mnras/stv1877}

\bibitem[{{Somerville} {et~al.}(2021){Somerville}, {Olsen}, {Yung}, {Pacifici},
  {Ferguson}, {Behroozi}, {Osborne}, {Wechsler}, {Pandya}, {Faber}, {Primack},
  \& {Dekel}}]{Somerville2021}
{Somerville}, R.~S., {Olsen}, C., {Yung}, L.~Y.~A., {et~al.} 2021, \mnras, 502,
  4858, \dodoi{10.1093/mnras/stab231}

\bibitem[{{Speagle} {et~al.}(2014){Speagle}, {Steinhardt}, {Capak}, \&
  {Silverman}}]{speagle}
{Speagle}, J.~S., {Steinhardt}, C.~L., {Capak}, P.~L., \& {Silverman}, J.~D.
  2014, The Astrophysical Journal Supplement Series, 214, 15,
  \dodoi{10.1088/0067-0049/214/2/15}

\bibitem[{{Steinhardt} {et~al.}(2016){Steinhardt}, {Capak}, {Masters}, \&
  {Speagle}}]{Steinhardt2016}
{Steinhardt}, C.~L., {Capak}, P., {Masters}, D., \& {Speagle}, J.~S. 2016,
  \apj, 824, 21, \dodoi{10.3847/0004-637X/824/1/21}

\bibitem[{{Stevans} {et~al.}(2021){Stevans}, {Finkelstein}, {Kawinwanichakij},
  {Wold}, {Papovich}, {Somerville}, {Yung}, {Sherman}, {Ciardullo}, {Dav{\'e}},
  {Florez}, {Gronwall}, \& {Jogee}}]{Stevans2021}
{Stevans}, M.~L., {Finkelstein}, S.~L., {Kawinwanichakij}, L., {et~al.} 2021,
  \apj, 921, 58, \dodoi{10.3847/1538-4357/ac0cf6}

\bibitem[{{Straatman} {et~al.}(2014){Straatman}, {Labb{\'e}}, {Spitler},
  {Allen}, {Altieri}, {Brammer}, {Dickinson}, {van Dokkum}, {Inami},
  {Glazebrook}, {Kacprzak}, {Kawinwanichakij}, {Kelson}, {McCarthy},
  {Mehrtens}, {Monson}, {Murphy}, {Papovich}, {Persson}, {Quadri}, {Rees},
  {Tomczak}, {Tran}, \& {Tilvi}}]{Straatman2014}
{Straatman}, C. M.~S., {Labb{\'e}}, I., {Spitler}, L.~R., {et~al.} 2014, \apjl,
  783, L14, \dodoi{10.1088/2041-8205/783/1/L14}

\bibitem[{{Straatman} {et~al.}(2016){Straatman}, {Spitler}, {Quadri},
  {Labb{\'e}}, {Glazebrook}, {Persson}, {Papovich}, {Tran}, {Brammer},
  {Cowley}, {Tomczak}, {Nanayakkara}, {Alcorn}, {Allen}, {Broussard}, {van
  Dokkum}, {Forrest}, {van Houdt}, {Kacprzak}, {Kawinwanichakij}, {Kelson},
  {Lee}, {McCarthy}, {Mehrtens}, {Monson}, {Murphy}, {Rees}, {Tilvi}, \&
  {Whitaker}}]{Straatman2016}
{Straatman}, C. M.~S., {Spitler}, L.~R., {Quadri}, R.~F., {et~al.} 2016, \apj,
  830, 51, \dodoi{10.3847/0004-637X/830/1/51}

\bibitem[{{Toft} {et~al.}(2014){Toft}, {Smol{\v{c}}i{\'c}}, {Magnelli},
  {Karim}, {Zirm}, {Michalowski}, {Capak}, {Sheth}, {Schawinski}, {Krogager},
  {Wuyts}, {Sanders}, {Man}, {Lutz}, {Staguhn}, {Berta}, {Mccracken}, {Krpan},
  \& {Riechers}}]{toft14}
{Toft}, S., {Smol{\v{c}}i{\'c}}, V., {Magnelli}, B., {et~al.} 2014, \apj, 782,
  68, \dodoi{10.1088/0004-637X/782/2/68}

\bibitem[{{Valentino} {et~al.}(2020){Valentino}, {Tanaka}, {Davidzon}, {Toft},
  {G{\'o}mez-Guijarro}, {Stockmann}, {Onodera}, {Brammer}, {Ceverino},
  {Faisst}, {Gallazzi}, {Hayward}, {Ilbert}, {Kubo}, {Magdis}, {Selsing},
  {Shimakawa}, {Sparre}, {Steinhardt}, {Yabe}, \& {Zabl}}]{Valentino2020}
{Valentino}, F., {Tanaka}, M., {Davidzon}, I., {et~al.} 2020, \apj, 889, 93,
  \dodoi{10.3847/1538-4357/ab64dc}

\bibitem[{{Valentino} {et~al.}(2023){Valentino}, {Brammer}, {Gould}, {Kokorev},
  {Fujimoto}, {Kragh Jespersen}, {Vijayan}, {Weaver}, {Ito}, {Tanaka},
  {Ilbert}, {Magdis}, {Whitaker}, {Faisst}, {Gallazzi}, {Gillman},
  {Gimenez-Arteaga}, {Gomez-Guijarro}, {Kubo}, {Heintz}, {Hirschmann}, {Oesch},
  {Onodera}, {Rizzo}, {Lee}, {Strait}, \& {Toft}}]{Valentino2023}
{Valentino}, F., {Brammer}, G., {Gould}, K. M.~L., {et~al.} 2023, arXiv
  e-prints, arXiv:2302.10936, \dodoi{10.48550/arXiv.2302.10936}

\bibitem[{{Vijayan} {et~al.}(2021){Vijayan}, {Lovell}, {Wilkins}, {Thomas},
  {Barnes}, {Irodotou}, {Kuusisto}, \& {Roper}}]{Vijayan2021}
{Vijayan}, A.~P., {Lovell}, C.~C., {Wilkins}, S.~M., {et~al.} 2021, \mnras,
  501, 3289, \dodoi{10.1093/mnras/staa3715}

\bibitem[{{Weaver} {et~al.}(2022){Weaver}, {Kauffmann}, {Ilbert}, {McCracken},
  {Moneti}, {Toft}, {Brammer}, {Shuntov}, {Davidzon}, {Hsieh}, {Laigle},
  {Anastasiou}, {Jespersen}, {Vinther}, {Capak}, {Casey}, {McPartland},
  {Milvang-Jensen}, {Mobasher}, {Sanders}, {Zalesky}, {Arnouts}, {Aussel},
  {Dunlop}, {Faisst}, {Franx}, {Furtak}, {Fynbo}, {Gould}, {Greve}, {Gwyn},
  {Kartaltepe}, {Kashino}, {Koekemoer}, {Kokorev}, {Le F{\`e}vre}, {Lilly},
  {Masters}, {Magdis}, {Mehta}, {Peng}, {Riechers}, {Salvato}, {Sawicki},
  {Scarlata}, {Scoville}, {Shirley}, {Silverman}, {Sneppen}, {Smolc̆i{\'c}},
  {Steinhardt}, {Stern}, {Tanaka}, {Taniguchi}, {Teplitz}, {Vaccari}, {Wang},
  \& {Zamorani}}]{Weaver2022}
{Weaver}, J.~R., {Kauffmann}, O.~B., {Ilbert}, O., {et~al.} 2022, \apjs, 258,
  11, \dodoi{10.3847/1538-4365/ac3078}

\bibitem[{{Whitaker} {et~al.}(2011){Whitaker}, {Labb{\'e}}, {van Dokkum},
  {Brammer}, {Kriek}, {Marchesini}, {Quadri}, {Franx}, {Muzzin}, {Williams},
  {Bezanson}, {Illingworth}, {Lee}, {Lundgren}, {Nelson}, {Rudnick}, {Tal}, \&
  {Wake}}]{Whitaker2011}
{Whitaker}, K.~E., {Labb{\'e}}, I., {van Dokkum}, P.~G., {et~al.} 2011, \apj,
  735, 86, \dodoi{10.1088/0004-637X/735/2/86}

\bibitem[{{Williams} {et~al.}(2018){Williams}, {Curtis-Lake}, {Hainline},
  {Chevallard}, {Robertson}, {Charlot}, {Endsley}, {Stark}, {Willmer},
  {Alberts}, {Amorin}, {Arribas}, {Baum}, {Bunker}, {Carniani}, {Crandall},
  {Egami}, {Eisenstein}, {Ferruit}, {Husemann}, {Maseda}, {Maiolino}, {Rawle},
  {Rieke}, {Smit}, {Tacchella}, \& {Willott}}]{Williams2018}
{Williams}, C.~C., {Curtis-Lake}, E., {Hainline}, K.~N., {et~al.} 2018, \apjs,
  236, 33, \dodoi{10.3847/1538-4365/aabcbb}

\bibitem[{{Williams} {et~al.}(2021){Williams}, {Oesch}, {Barrufet}, {Bezanson},
  {Bowler}, {Brammer}, {Dayal}, {Franx}, {Hutter}, {Labbe}, {Maseda}, {Ucci},
  \& {Whitaker}}]{Williams2021}
{Williams}, C.~C., {Oesch}, P., {Barrufet}, L., {et~al.} 2021, {PANORAMIC - A
  Pure Parallel Wide Area Legacy Imaging Survey at 1-5 Micron}, JWST Proposal.
  Cycle 1, ID. \#2514

\bibitem[{{Williams} {et~al.}(2009){Williams}, {Quadri}, {Franx}, {van Dokkum},
  \& {Labb{\'e}}}]{Williams2009}
{Williams}, R.~J., {Quadri}, R.~F., {Franx}, M., {van Dokkum}, P., \&
  {Labb{\'e}}, I. 2009, \apj, 691, 1879, \dodoi{10.1088/0004-637X/691/2/1879}

\bibitem[{{Yung} {et~al.}(2019){Yung}, {Somerville}, {Finkelstein}, {Popping},
  \& {Dav{\'e}}}]{Yung2019a}
{Yung}, L.~Y.~A., {Somerville}, R.~S., {Finkelstein}, S.~L., {Popping}, G., \&
  {Dav{\'e}}, R. 2019, \mnras, 483, 2983, \dodoi{10.1093/mnras/sty3241}

\bibitem[{{Yung} {et~al.}(2022){Yung}, {Somerville}, {Ferguson}, {Finkelstein},
  {Gardner}, {Dav{\'e}}, {Bagley}, {Popping}, \& {Behroozi}}]{Yung2022}
{Yung}, L.~Y.~A., {Somerville}, R.~S., {Ferguson}, H.~C., {et~al.} 2022,
  \mnras, 515, 5416, \dodoi{10.1093/mnras/stac2139}

\bibitem[{{Zavala} {et~al.}(2017){Zavala}, {Aretxaga}, {Geach}, {Hughes},
  {Birkinshaw}, {Chapin}, {Chapman}, {Chen}, {Clements}, {Dunlop}, {Farrah},
  {Ivison}, {Jenness}, {Micha{\l}owski}, {Robson}, {Scott}, {Simpson},
  {Spaans}, \& {van der Werf}}]{Zavala2017}
{Zavala}, J.~A., {Aretxaga}, I., {Geach}, J.~E., {et~al.} 2017, \mnras, 464,
  3369, \dodoi{10.1093/mnras/stw2630}

\bibitem[{{Zavala} {et~al.}(2018){Zavala}, {Aretxaga}, {Dunlop},
  {Micha{\l}owski}, {Hughes}, {Bourne}, {Chapin}, {Cowley}, {Farrah}, {Lacey},
  {Targett}, \& {van der Werf}}]{Zavala2018ceers}
{Zavala}, J.~A., {Aretxaga}, I., {Dunlop}, J.~S., {et~al.} 2018, \mnras, 475,
  5585, \dodoi{10.1093/mnras/sty217}

\bibitem[{{Zavala} {et~al.}(2022){Zavala}, {Casey}, {Spilker}, {Tadaki},
  {Tsujita}, {Champagne}, {Iono}, {Kohno}, {Manning}, \&
  {Monta{\~n}a}}]{Zavala2022}
{Zavala}, J.~A., {Casey}, C.~M., {Spilker}, J., {et~al.} 2022, \apj, 933, 242,
  \dodoi{10.3847/1538-4357/ac7560}

\bibitem[{{Zavala} {et~al.}(2023){Zavala}, {Buat}, {Casey}, {Finkelstein},
  {Burgarella}, {Bagley}, {Ciesla}, {Daddi}, {Dickinson}, {Ferguson}, {Franco},
  {Jim{\'e}nez-Andrade}, {Kartaltepe}, {Koekemoer}, {Bail}, {Murphy},
  {Papovich}, {Tacchella}, {Wilkins}, {Aretxaga}, {Behroozi}, {Champagne},
  {Fontana}, {Giavalisco}, {Grazian}, {Grogin}, {Kewley}, {Kocevski},
  {Kirkpatrick}, {Lotz}, {Pentericci}, {P{\'e}rez-Gonz{\'a}lez}, {Pirzkal},
  {Ravindranath}, {Somerville}, {Trump}, {Yang}, {Yung}, {Almaini},
  {Amor{\'\i}n}, {Annunziatella}, {Haro}, {Backhaus}, {Barro}, {Bell},
  {Bhatawdekar}, {Bisigello}, {Buitrago}, {Calabr{\`o}}, {Castellano},
  {Ch{\'a}vez Ortiz}, {Chworowsky}, {Cleri}, {Cohen}, {Cole}, {Cooke},
  {Cooper}, {Cooray}, {Costantin}, {Cox}, {Croton}, {Dav{\'e}}, {de La Vega},
  {Dekel}, {Elbaz}, {Estrada-Carpenter}, {Fern{\'a}ndez}, {Finkelstein},
  {Freundlich}, {Fujimoto}, {Garc{\'\i}a-Argum{\'a}nez}, {Gardner}, {Gawiser},
  {G{\'o}mez-Guijarro}, {Guo}, {Hamilton}, {Hathi}, {Holwerda}, {Hirschmann},
  {Huertas-Company}, {Hutchison}, {Iyer}, {Jaskot}, {Jha}, {Jogee}, {Juneau},
  {Jung}, {Kassin}, {Kurczynski}, {Larson}, {Leung}, {Long}, {Lucas},
  {Magnelli}, {Mantha}, {Matharu}, {McGrath}, {McIntosh}, {Medrano}, {Merlin},
  {Mobasher}, {Morales}, {Newman}, {Nicholls}, {Pandya}, {Rafelski}, {Ronayne},
  {Rose}, {Ryan}, {Santini}, {Seill{\'e}}, {Shah}, {Shen}, {Simons}, {Snyder},
  {Stanway}, {Straughn}, {Teplitz}, {Vanderhoof}, {Vega-Ferrero}, {Wang},
  {Weiner}, {Willmer}, {Wuyts}, \& {(The Ceers Team)}}]{Zavala2023}
{Zavala}, J.~A., {Buat}, V., {Casey}, C.~M., {et~al.} 2023, \apjl, 943, L9,
  \dodoi{10.3847/2041-8213/acacfe}

\end{thebibliography}
\bibliographystyle{aasjournal}

\appendix
\restartappendixnumbering

In addition to the two wedges presented in this work, we also explored a wider wedge. This ``red selection wedge'' was explored as an option to capture potentially dust reddened quiescent galaxies (see e.g. the BEAGLE models presented in Figure \ref{fig:sim_color_selection}). The red selection wedge has the same SNR requirements and uses the same three bands as the other wedges, but instead requires that galaxies meet the following criteria: 

\begin{multline}
    \hspace{-0.4cm}\mathrm{Red-A.}\indent(F150W - F277W) < 1.5 + 6.25 \times (F277W-F444W) \\
    \mathrm{and} \\
    \mathrm{Red-B.}\indent (F150W - F277W) > 1.15 - 0.5 \times (F277W-F444W)\\
    \mathrm{and} \\
    \mathrm{Red-C.}\indent (F150W - F277W) > -0.6 + 2 \times (F277W-F444W)
\end{multline} \label{eqn:red}

The red selection wedge has the same Criterion A as the first two wedges presented in this work, the same Criterion B as the long wedge, and a new Criterion C which reaches redder into the $F277W-F444W$ color space. The expansion of this wedge space is represented by the red dashed lines in Appendix Figures \ref{fig:ceers_wide_qgs}$-$\ref{fig:color_selection_appendix}. 

Only seven additional $z\gtrsim3$ quiescent galaxy candidates were recovered in this red wedge space, while the number of contaminants ballooned to an additional $\sim$500 sources, demonstrating the diminishing returns in widening the wedge to capture more red sources. As expected based on the empirically-constrained SEDs in Figure \ref{fig:seds} of the main text, the majority of these contaminants are fit as moderately to heavily obscured (A$_V \gtrsim2$) star forming galaxies at intermediate redshifts ($z\sim1-4$). See Figure \ref{fig:ceers_interlopers} below.

We also show this wide red wedge applied to simulations in Figure \ref{fig:color_selection_appendix}. The Santa Cruz SAM appears to model a significant population of red objects in this wider red wedge space that looks similar to the observations as shown in Figure \ref{fig:ceers_wide_qgs}, but none are quiescent galaxies at $z\ge3$. However, the \textsc{eagle} and \textsc{flares} simulations predict a population of quiescent galaxies at $z=3$, 4, and 6 that would be best captured by the red wedge. Future discoveries, if any, of massive $z>5$ quiescent galaxies will be useful in testing these color predictions. 

\begin{figure*}[ht!]
\begin{center}
\includegraphics[trim=1cm 1cm 1cm 0cm, width=0.95\textwidth]{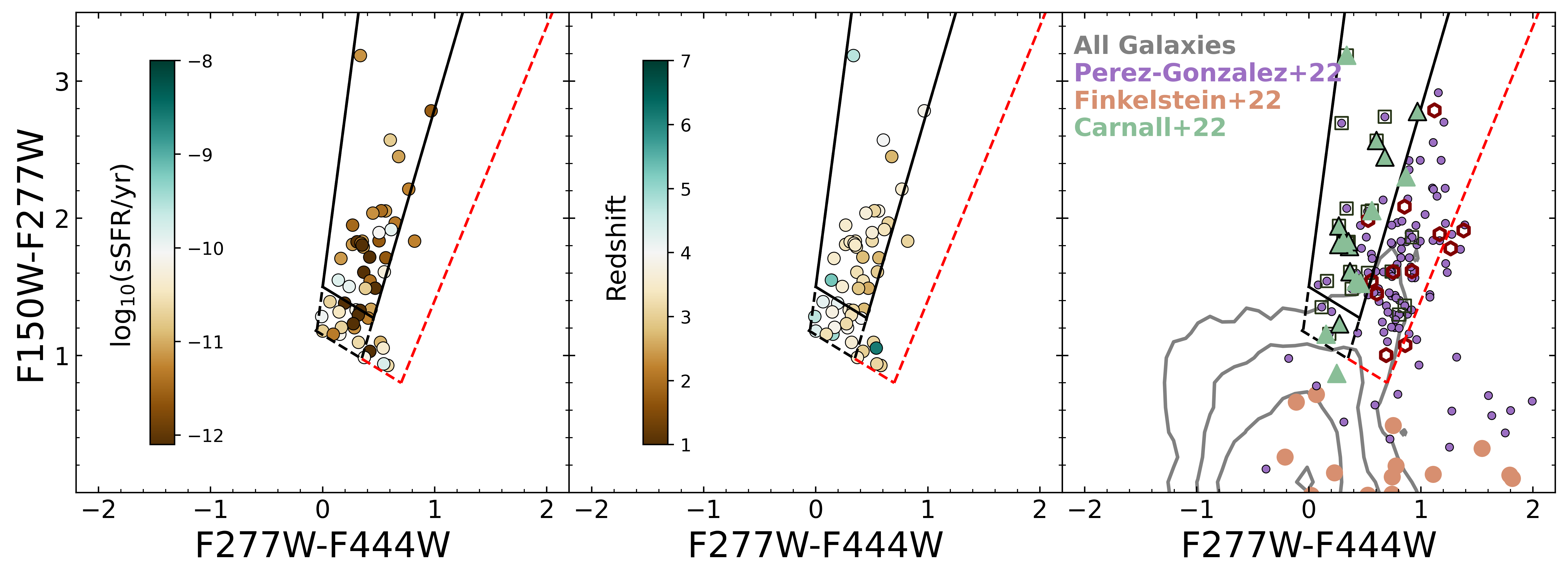}
\end{center}
\caption{Same as Figure \ref{fig:ceers} in the main text, but here we show the additional handful ($n = 7$) of candidate $z\gtrsim3$ quiescent galaxies recovered in the widened red wedge space. As shown in the following figure, the red wedge is heavily contaminated by dust obscured galaxies, with diminishing returns on the identification of quiescent galaxies at high-$z$.} \label{fig:ceers_wide_qgs}
\end{figure*}

\begin{figure*}[ht!]
\begin{center}
\includegraphics[trim=1cm 1cm 1cm 0cm, width=0.95\textwidth]{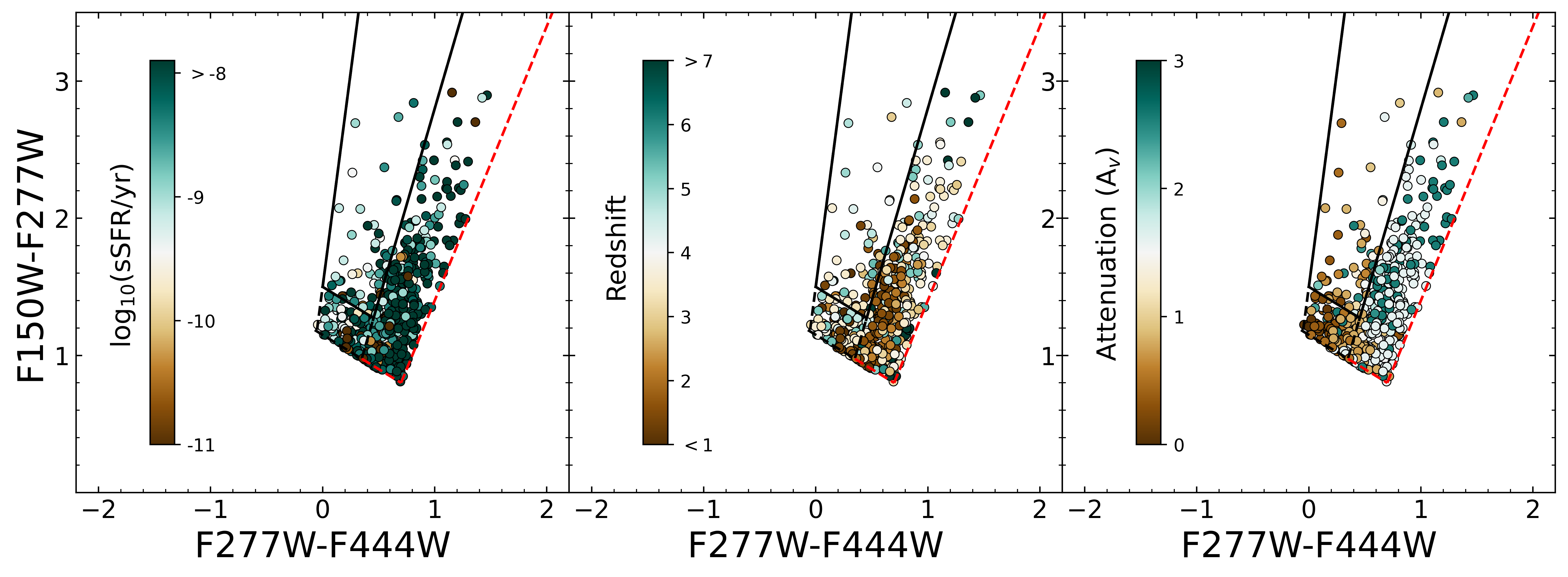}
\end{center}
\caption{Same as previous figure except here we show only the contaminants in each wedge. As expected from SED templates and existing literature studies in CEERS (e.g. \citealt{Perez-Gonzalez2022}), the wide/red wedge is occupied almost entirely by heavily dust obscured galaxies with A$_\mathrm{V} \ge 1$ at Cosmic Noon ($z=1-4$). } \label{fig:ceers_interlopers}
\end{figure*}

\begin{figure*}[ht!]
\begin{center}
\includegraphics[trim=0cm 1cm 0cm 0cm, width=1\textwidth]{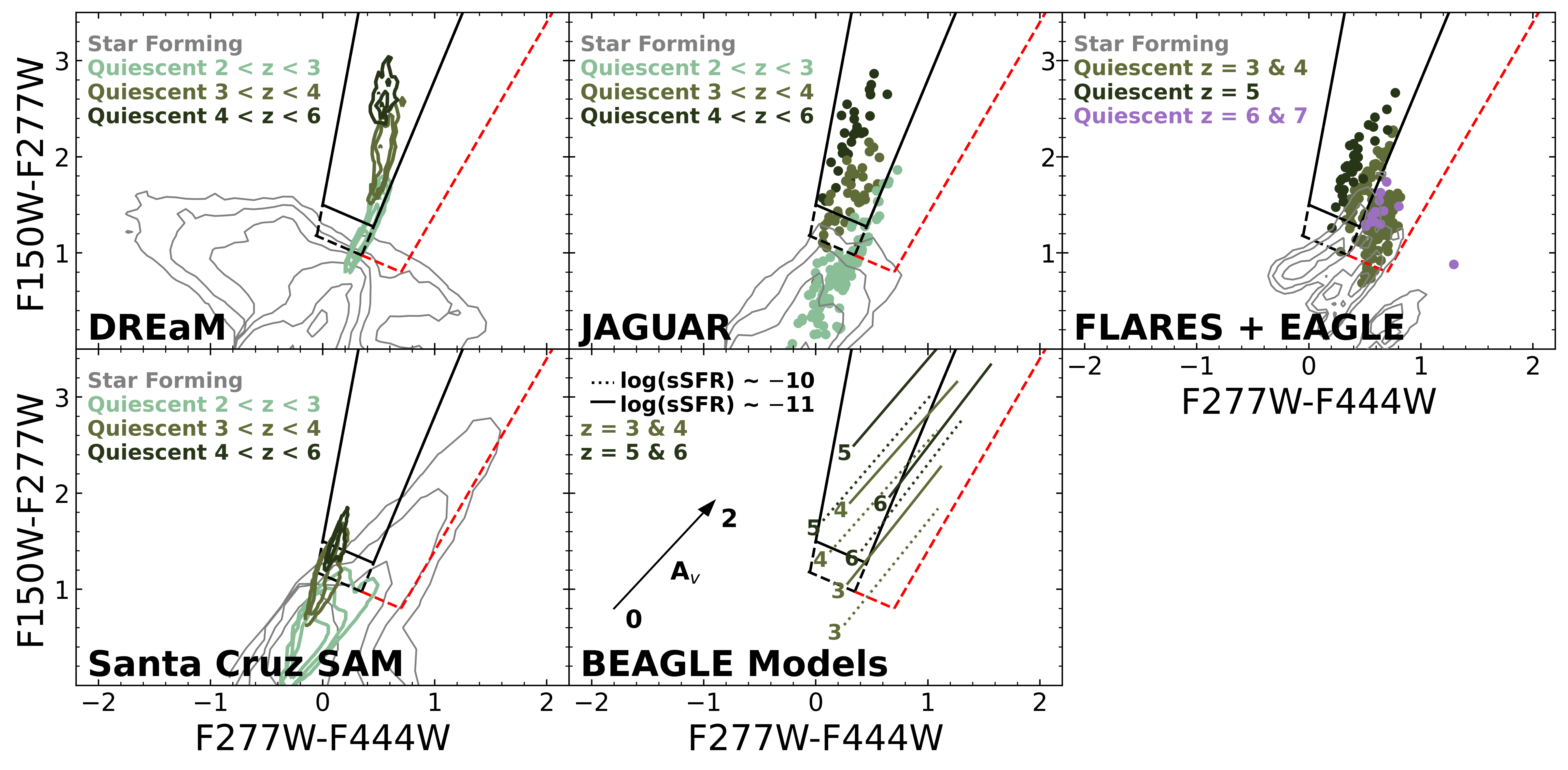}
\end{center}
\caption{Same as Figure \ref{fig:seds} and \ref{fig:sim_color_selection} except with with red wedge option overlaid in red dashed lines. The red wedge appears to capture $z\ge3$ quiescent galaxies primarily in the \textsc{eagle} and \textsc{flares} simulations. } \label{fig:color_selection_appendix}
\end{figure*}

\end{document}